\documentclass[preprint,amsmath,amssymb,superscriptaddress,showpacs]{revtex4}
\usepackage{graphicx}
\usepackage{dcolumn}
\usepackage{bm}
\usepackage{stmaryrd}
\usepackage{latexsym}
\usepackage{amssymb}
\usepackage{amsfonts}
\usepackage{amsmath}
\usepackage{fancybox}
\usepackage{color}

\definecolor{MyDarkBlue}{rgb}{0,0.08,0.45}

\begin{document}
\title{The effect of sublattice symmetry breaking on the electronic properties of a doped graphene}
\author{Alireza Qaiumzadeh}
\address {School of Physics, Institute for Research in
Fundamental Sciences (IPM), Tehran 19395-5531, Iran}
\address {Institute for Advanced Studies in Basic Sciences (IASBS),
Zanjan, 45195-1159, Iran}
\author{Reza Asgari~\footnote{Corresponding author: Tel: +98 21 22280692; fax: +98 21 22280415.\\ E-mail address: asgari@theory.ipm.ac.ir, }}
\address {School of Physics, Institute for Research in
Fundamental Sciences (IPM), Tehran 19395-5531, Iran}
\begin{abstract}
Motivated by a number of recent experimental studies, we have carried out the microscopic calculation of the quasiparticle self-energy and spectral function in a doped graphene
when a symmetry breaking of the sublattices is occurred. Our systematic
study is based on the many-body G$_0$W approach that is established on the random phase approximation and on graphene's massive Dirac equation continuum model. We report extensive calculations of both the
real and imaginary parts of the quasiparticle self-energy in the presence of a gap opening.
We also present results for spectral function, renormalized Fermi
velocity and band gap renormalization of massive Dirac Fermions over a broad
range of electron densities.
We further show that the mass generating in graphene
washes out the plasmaron peak in spectral weight.

\end{abstract}
\maketitle

\section{INTRODUCTION}

Graphene is a single atomic layer of crystalline carbon on the
honeycomb lattice consists of two interpenetrating
triangular sublattices A and B, has opened up a new field for
fundamental studies and applications.~\cite{novoselov1,novoselov2,novoselov3,novoselov4} Peculiar
electronic properties of graphene give rise possibility to come over
silicon-based electronics limitations.~\cite{carbon} The single-particle energy
spectrum in graphene contains two zero-energy at $K^+$ and $K^-$
points of the Brillouin zone which are called as valleys or Dirac
points. Due to the presence of the two carbon atoms per unit cell, the
quasiparticle (QP) need to be described by a two component wave function.

The charge carriers in a pristine graphene
show linear and isotropic energy dispersion relation and massless
chiral behavior for the energy scales up to 1 eV. Recently, graphene has revealed a variety of unusual
transport phenomena characteristics of two-dimensional (2D) Dirac Fermions such as an anomalous
integer quantum Hall effect at room temperature, a minimum quantum conductivity, Klein tunneling paradox, weak and
anti-localization, an absence of Wigner crystallization phase and Shubnikov-deHaas oscillations that exhibit a phase shift of $\pi$ due to Berry's phase.~\cite{graphene_rev1,graphene_rev2,graphene_rev3,graphene_rev4,graphene_rev5,graphene_rev6,tomadin}
One important difference between conventional electron gas and Dirac Fermion particle is that the contribution of exchange and correlation to the chemical potential is
an increasing rather than a decreasing function of carrier-density. This property
implies that exchange and correlation increase the effectiveness of screening, in contrast to the usual case in which exchange and correlation weakens screening. This unusual property follows from the difference in sublattice pseudospin
chirality between the Dirac model's negative energy valence band states and its conduction band states.

The massless Dirac-like carriers in graphene
have almost semi-ballistic transport behavior with small resistance
due to the suppression of back-scattering process, and moreover graphene
is a good thermal conductor.~\cite{thermal} The mobility of carriers in
graphene is quite high~\cite{morozov1,morozov2,morozov3,morozov4} which is much higher than the electron
mobility revealed on the semiconductor hetrostructures.~\cite{eng1,eng2}
On the other hand, by measuring the stiffness of materials it is shown that graphene is the strongest
material in two-dimension structures.~\cite{strong1,strong2} These properties as well as
capability to control of the type and density of charge carriers by
gate voltage or the chemical doping~\cite{dop1,dop2,dop_lanzara1,dop_lanzara2} make
graphene an ideal candidate for superior nano-electronic devices
operating at high frequencies.

Most electronic applications are based on the presence of a gap
between the valence and conduction bands in the conventional
semiconductors. The band gap is a measure of the threshold voltage
and on-off ratio of the field effect transistors (FETs).~\cite{FET1,FET2}
Therefore, for integrating graphene into semiconductor technology, it is
crucial to induce a band gap in Dirac points to control the
transport of carriers. Consequently, band gap engineering in graphene
is a hot topic with fundamental and applied
significance.~\cite{gap} In the literature several routes have being
proposed and applied to induce and control a gap in graphene. One of them is using quantum
confined geometries such as quantum dots and
nanoribbons.~\cite{gap_ribbon1,gap_ribbon2,gap_ribbon3,gap_ribbon4,gap_ribbon5} It is shown that the gap values
increases by decreasing of nanoribbon width. Another alternative way is
spin-orbit coupling whose origin is due to both intrinsic spin-orbit interactions and the Rashba
interaction.~\cite{gap_spin1,gap_spin2,gap_spin3,gap_spin4} Another method to generate a gap in
graphene sheets is an inversion symmetry breaking of the sublattices when the number of electrons
on A and B atoms are different~\cite{gapsub1,gapsub2,gapsub3,gapsub4} or Kekul\'{e}~\cite{kekule1} distortion,
\textit{e.g.} graphene on proper
substrates~\cite{dop_lanzara1,dop_lanzara2,lanzara1,lanzara2,eva,gruneis1,gruneis2,giovannetti} or
adsorb of some molecules such as water, ammonia~\cite{ribeiro1,ribeiro2} and
CrO$_3$~\cite{zanella} or an alkali-metal sub-monolayer on graphene
sheets.

Recently angle resolved photoemission spectroscopy ( ARPES) experiments on graphene epitaxially grown on SiC and \textit{ab initio}
simulations reported a gap opening in the band structure of graphene placed
on proper substrates, and suggested that interactions
between the graphene sheet and the substrate leads to symmetry breaking of the
A and B sublattices and it consequences to induce a gap in the band
structure. Experimenters~\cite{dop_lanzara1,dop_lanzara2,lanzara1,lanzara2,kruczynski} observed
a gap of 260 meV in band structure of the epitaxial graphene on
SiC substrate due to interaction with substrate. In addition, Zhou {\it et al.}~\cite{dop_lanzara1} found a reversible metal-insulator transition and a fine tuning of the
carriers from electron to hole by molecular doping in gapped graphene. A Density Functional Theory
(DFT) calculation confirmed a substrate induced symmetry breaking.~\cite{kim}. Their results showed a gap in the band spectra of graphene about 200 meV which is in
agreement with recent experimental observation. Their
calculation determined that there is a 140 meV on-site energy difference between two sublattices. In addition, a band gap is observed in
spectra of graphene on Ni(111) substrate~\cite{gruneis1,gruneis2} as well as
a gap about of 10 meV in suspended graphene above a graphite
substrate~\cite{eva} due to sublattice symmetry breaking mechanism.
Moreover, based on the \textit{ab initio} calculations, it is
suggested that boron nitride substrate induced a gap of
53 meV.~\cite{giovannetti} Note that the gap value calculated within DFT is in general underestimating the true band gap value.

In this paper we consider the sublattice symmetry breaking mechanism for a gap opening in a pristine doped graphene sheet and
study the impact of gap upon some electronic properties of
QPs. To investigate the influence
of gap in the many-body properties of QP in graphene we
use the random phase approximation (RPA) and the G$_0$W
approximation. It should be noted that a detailed analysis provided a framework for the microscopic evaluation of the QP-QP interaction in the {\it gapless graphene} by means of the RPA was carried out by us in Ref.~[\onlinecite{im1}] At the beginning, we review briefly the results of the ground state thermodynamic properties that we have already presented elsewhere.~\cite{alireza} Our new results are based on the QP
self-energy properties in the presence of a gap opening in the electronic spectrum. From the self-energy we then obtain the QP energies, renormalized Fermi velocity, spectral function which can be compared with ARPES spectra and finally the band gap renormalization of massive Dirac Fermions in doped graphene.
We have shown that mass generating in graphene washes out a satellite band in the spectral function in agreement with recent experimental observations.~\cite{lanzara1}

This paper is organized as followed. In Section II we
introduce our model Hamiltonian and then review some ground state
properties of gapped graphene. In Section III we focus on the
properties of imaginary and real parts of self-
energy for gapped graphene and then calculate QP spectral function,
renormalized Fermi velocity and band gap renormalization. Finally we conclude in Section IV.

\section{GROUND STATE THERMODYNAMIC PROPERTIES}

We consider the sublattice symmetry breaking mechanism in which the densities of particles associated to on-site energy
$\mu_{a(b)}$, for A(B) sublattice are different. The electronic structure of graphene can be reasonably good described using a rather simple tight-binding Hamiltonian, leading to analytical solutions for their energy dispersion and related eigenstates. The noninteracting
tight binding Hamiltonian for $\pi$ band electrons is determined by~\cite{gapsub1,gapsub2,gapsub3,gapsub4}
\begin{eqnarray}
\hat{H_0}&=&t\sum_{i}(a_i^\dagger
b_i+{\rm c.c.})+\mu_a\sum_{i}a^\dagger_ia_i+\mu_b\sum_{i}b^\dagger_ib_i\nonumber\\
&=&t\sum_{i}(a_i^\dagger b_i+{\rm
c.c.})+\frac{\mu_a-\mu_b}{2}\sum_{i}(a^\dagger_ia_i-b^\dagger_ib_i)
+\frac{\mu_a+\mu_b}{2}\sum_{i}(a^\dagger_ia_i+b^\dagger_ib_i)
\end{eqnarray}
where the sums run over unit cells, $t\simeq2.7$ eV denotes the nearest
neighbor hopping parameter and $a_i(b_i)$ is Fermi annihilation operator acts on
A(B) sublattice. The second term in the noninteracting Hamiltonian
breaks the inversion symmetry and causes to a band gap with value of
$2\Delta=|\mu_a-\mu_b|$ at the Dirac points. The last term is a constant and we left it out. The effective
Hamiltonian at low excited energies lead to a
2D massive Dirac Hamiltonian, $\mathcal{\hat{H}}_0=\hbar
v_{\rm F}{\vec{\sigma}}\cdot{\bf k}+\Delta\sigma_3$, where $\vec{\sigma}$ are Pauli matrices and $v_{\rm F}=3ta/2\hbar\simeq10^6$ m/s is the Fermi
velocity where $a\simeq1.42$ {\AA} is the carbon-carbon
distance in honeycomb lattice. The two eigenvalues of noninteracting Hamiltonian are given by
$E_{\bf k}=\pm\sqrt{(\hbar v_{\rm F}k)^2+\Delta^2}$ for conduction band (+) and
valance band (-) which is a fully occupied. In addition, the model Hamiltonian can be used as an
approximated model for describing a graphene antidot lattice in the
vicinity of a band gap with a small effective mass
value~\cite{antidot}, or moreover used as an effective Hamiltonian for the intrinsic
spin-orbit interaction in graphene where $\Delta=\Delta_{\rm SO}$
is the strength of the spin-orbit interaction.~\cite{gap_spin1,gap_spin2,gap_spin3,gap_spin4} If
$\mu_a=\mu_b$ the Hamiltonian reduces to massless Dirac
Hamiltonian with two chiral eigenstates having
the conical band structures $\varepsilon_k=\pm\hbar
v_{\rm F}k$.

We consider the long-range Coulomb electron-electron interaction.
We left out the intervalley scattering and use the two
component Dirac Fermion model. Accordingly, the total interacting Hamiltonian in
a continuum model at $K^+$ point is expressed as
\cite{yafis,Giuliani}
\begin{equation}\label{ham}
\hat{H}= \sum_{{\bf k},\sigma}\Psi^\dagger_{{\bf
k},\sigma}\mathcal{\hat{H}}_0\Psi_{{\bf k},\sigma}+
\frac{1}{2S}\sum_{{\bf q}\neq 0}V_q ({\hat n}_{\bf q} {\hat
n}_{-{\bf q}}-{\hat N}),
\end{equation}
where $\Psi^\dagger_{{\bf k},\sigma}=(\psi^a_{+,\sigma}({\bf
k}),\psi^b_{+,\sigma}({\bf k}))$ is two component pseudospinors of
the noninteracting Hamiltonian, $S$ is the sample area, ${\hat N}$
is the total number operator and $V_q=2\pi e^2/\epsilon q$ is the
bare Coulomb interaction where $\epsilon$ is an average dielectric constant of the
surrounding medium. The coupling constant in graphene
is $\alpha_{gr}=g_sg_ve^2/\epsilon\hbar v_{\rm F}$ where $g_s=g_v=2$ being
the spin and valley degeneracy, respectively. The coupling
constant in graphene depends only on the substrate dielectric
constant while in the conventional 2D electron systems is density
dependent. The typical value of dimensionless coupling constant is 1 or 2
for graphene supported on a substrate such a SiC or SiO$_2$.

A central quantity in the many-body techniques is the
noninteracting dynamical polarizability function $\chi^{(0)}({\bf
q},i\omega,\mu)$ where $\mu$ is the chemical potential. The problem of linear density response
is set up by considering a fluid described by the Hamiltonian, $\hat{H}$, which is subject to an external potential. The external potential must be sufficiently weak for low-order perturbation theory to suffice. The induced density change has a linear relation to the external potential through the noninteracting dynamical polarizability function. This
function is recently calculated along the imaginary frequency
axis and it is given by~\cite{alireza}
\begin{eqnarray}\label{eq:final_result}
\chi^{(0)}({\bf q},i \omega,\mu\geq\Delta)&=&-\frac{g_sg_v}{2\pi
\hbar^2v_{\rm F}^2}\{\mu- \Delta+\frac{
\varepsilon_q^2}{2}(\frac{\Delta}{{y^2}}+\frac{x_-^2}{2y}
\tan^{-1}(\frac{y} {2\Delta}))\nonumber\\&-&
\frac{\varepsilon_q^2}{4y}\Re
e\left[x_-^2(\sin^{-1}\frac{z(\mu)}{x_+}-\sin^{-1}\frac{z(\Delta)}{x_+})\right]\nonumber\\&+&
\frac{\varepsilon_q^2}{4y}\Re
e\left[z(\mu)\sqrt{x_+^2
-z^2(\mu)}-z(\Delta)\sqrt{x_+^2 -z^2(\Delta)}\right]\},
\end{eqnarray}
where
$x_{\pm}=\sqrt{1\pm4\Delta^2/(\varepsilon_q^2+\hbar^2\omega^2)}$,
$y=\sqrt{\varepsilon_q^2+\hbar^2 \omega^2}$ and
$z(x)=(2x+i\hbar\omega)/\varepsilon_q$. The Fermi energy
of a 2D massive Dirac Fermion system is given by
$E_{\rm F}=\mu=\sqrt{(\hbar v_{\rm F}k_{\rm F})^2+\Delta^2}$ and the Fermi
wavevector is $k_{\rm F}=\sqrt{4\pi n/g_sg_v}$ where $n$ is the density
of carriers. The noninteracting density of states (DOS) is determined by
$D(E)=g_sg_v|E|/2\pi\hbar^2v_{\rm F}^2\Theta(E^2-\Delta^2)$ which is density dependent at the Fermi surface.
It should be noticed that $D(E_{\rm F})$ equals to $m/2\pi\hbar^2$
in the conventional 2D electron gas system.
Here, $\Theta(x)$ is Heaviside step function.


We now turn to present our first numerical results which are based on the noninteracting polarization function.
The static polarization function as a function of
wavevector for various gap values is shown in Fig.~1(a). The static
polarization function in gapless case is a smooth function whereas a kink at $q=2k_{\rm F}$ occurs for gapped graphene and thus
the derivatives of $\chi^{(0)}({\bf q},0,\Delta\neq0)$ has a singular feature.
The singular behavior is the source of several phenomena such as the
Friedel oscillations and moreover the Ruderman-Kittel-Kasuya-Yoshida (RKKY)
interaction which the later is absent in gapless graphene. In Figs.~1(b)
and (c) we have plotted the dynamic polarization function as a function of frequency for wave
vectors smaller and larger than $q=2k_{\rm F}$, respectively. $\chi^{(0)}({\bf q},i\omega)$ tends to zero like $\omega^{-1}$ at large frequency region.

The polarization function along the real $\omega$
axis can be obtained by performing
analytical continuation of Eq.~(\ref{eq:final_result}).~\cite{alireza,pyatkovskiy1,pyatkovskiy2}
In Fig.~2 we have presented the real and imaginary parts of the
noninteracting polarization function as a function of frequency. Sharp cutoffs in the imaginary part of
$\chi^{(0)}({\bf q},\omega)$ are related to
the rapid swing in the real part of $\chi^{(0)}({\bf q},\omega)$. These
behaviors are in result of the fact that the real and imaginary
parts of the polarization function are related through the Kramers-Kr\"{o}nig
relations. Importantly, the sign change of the real part from negative to
positive shows a sweep across the electron-hole continuum. At very large gap values, the polarization function of
massive Dirac Fermions can be reduced to the polarization function (the
Lindhard's function) of conventional two dimensional electron gas
systems, as they are determined in Figs.~1 and 2.
Consequently, we settle
under situation that we can describe a range of band structures
from the Dirac's cone (gapless graphene) to the parabolic
(conventional semiconductors) band structure behavior by tuning the gap values from zero to a large value, respectively. We limited our calculations to the intermediate values of of $\Delta$ and we thus
expect wide range of the particular properties related to unique behavior of the polarization function.

\begin{figure}[t]
\includegraphics[width=5cm]{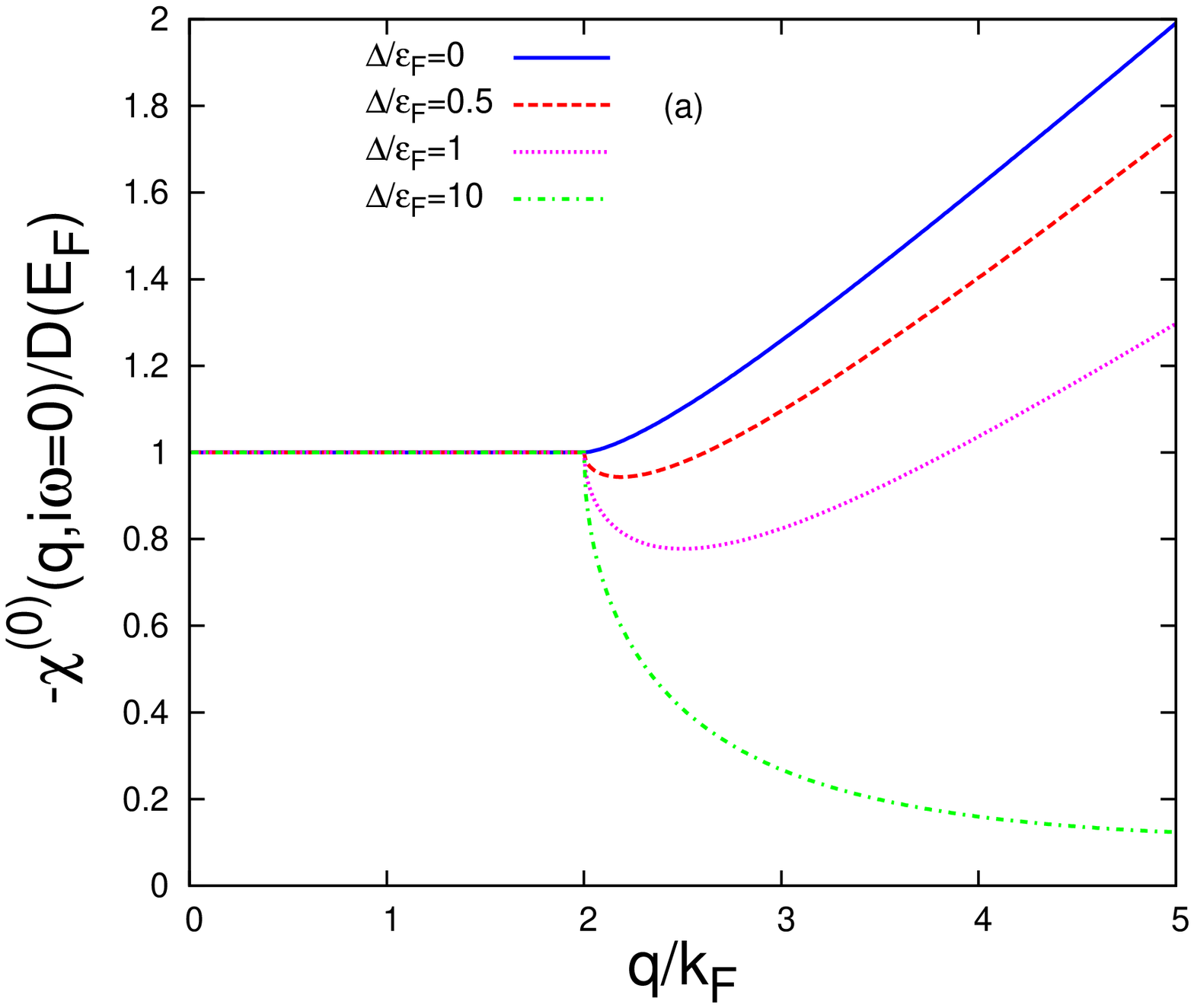}
\includegraphics[width=5cm]{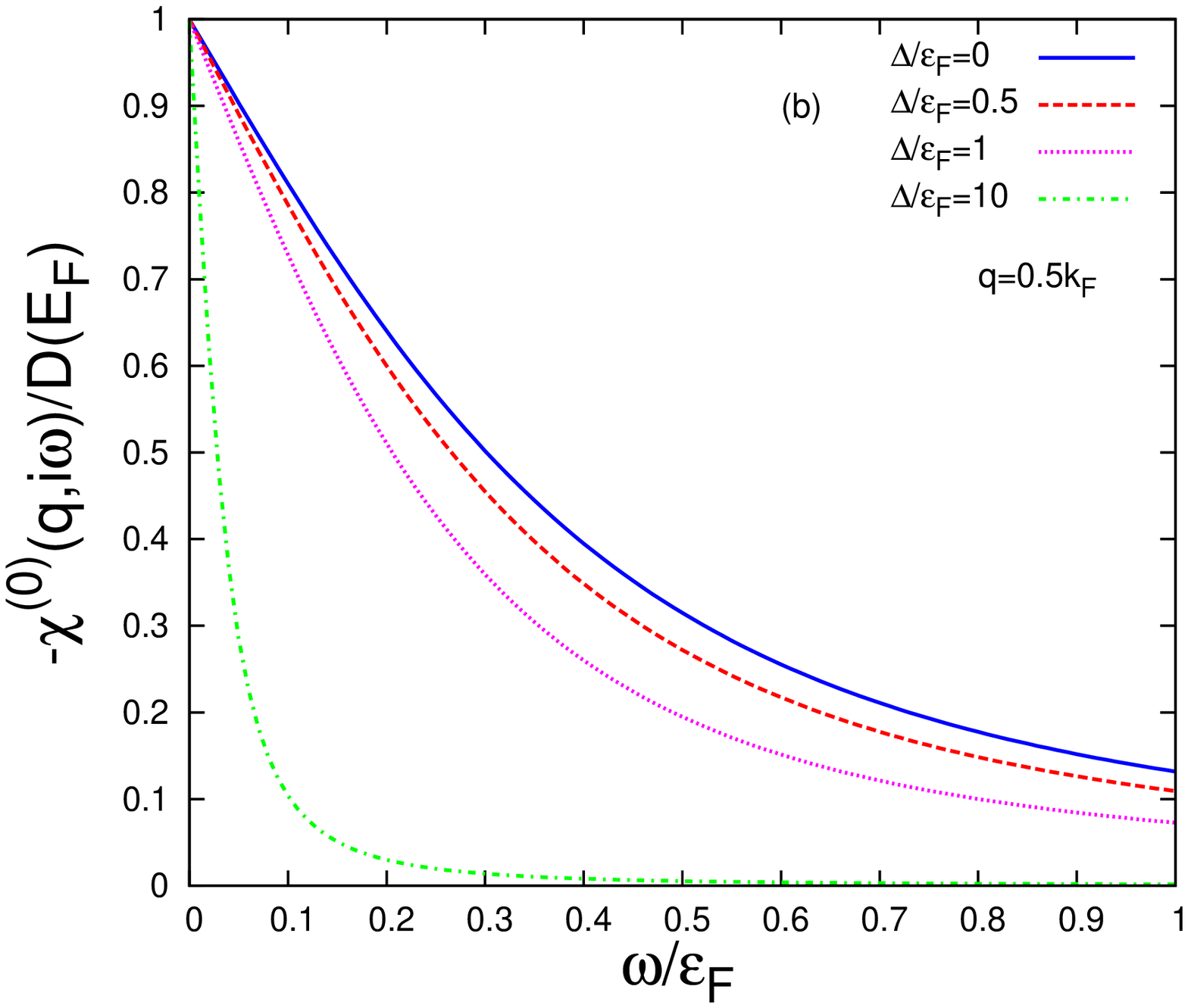}
\includegraphics[width=5cm]{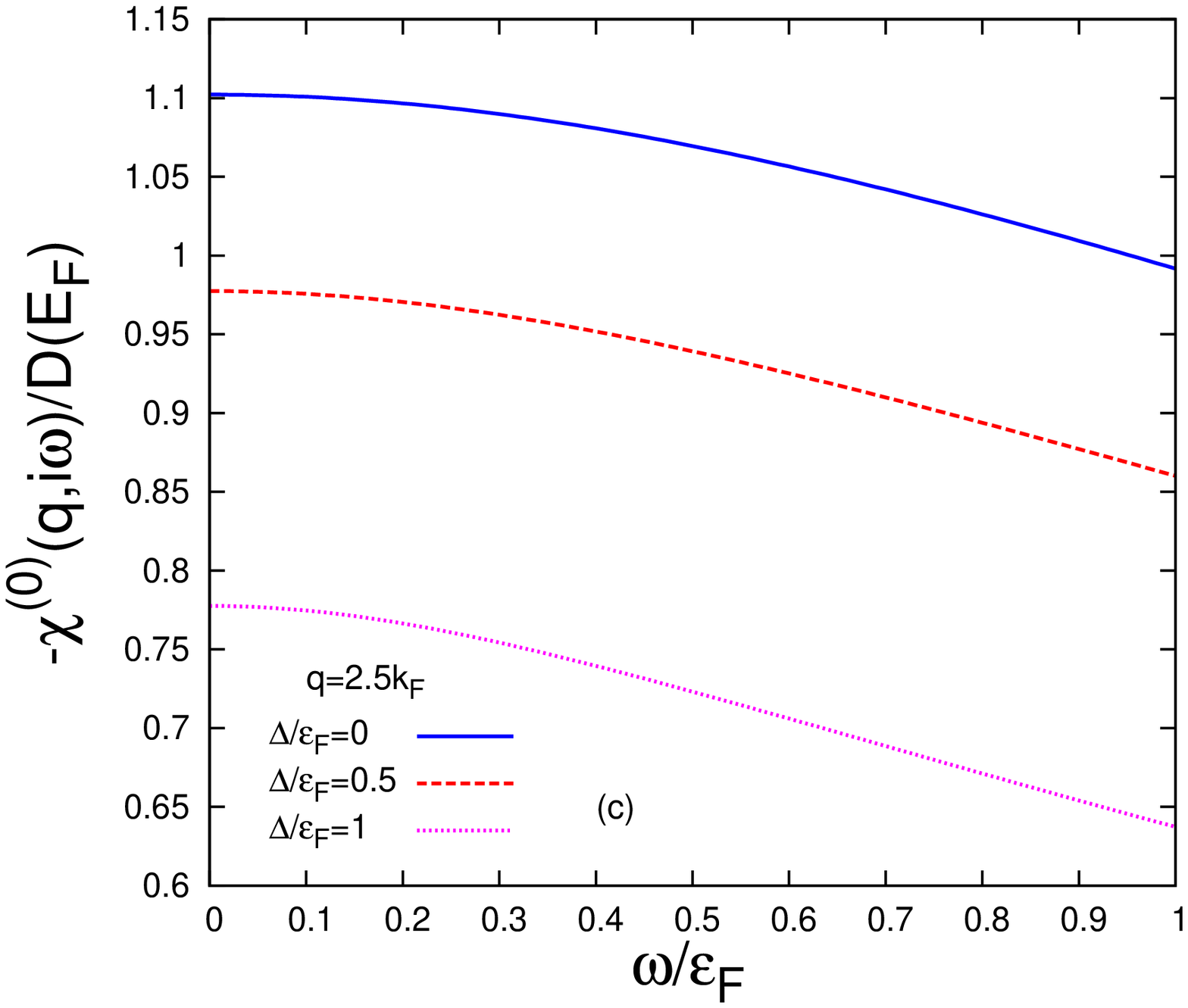}
\caption{(Color online) (a): The static noninteracting
polarization function as a function of $q$ for various $\Delta$.
The dimensionless noninteracting dynamic polarization at (b): $q=0.5 k_{\rm F}$
and (c): $q=2k_{\rm F}$ as a function of $\omega$ for various
$\Delta$.}
\end{figure}

\begin{figure}[b]
\includegraphics[width=5cm]{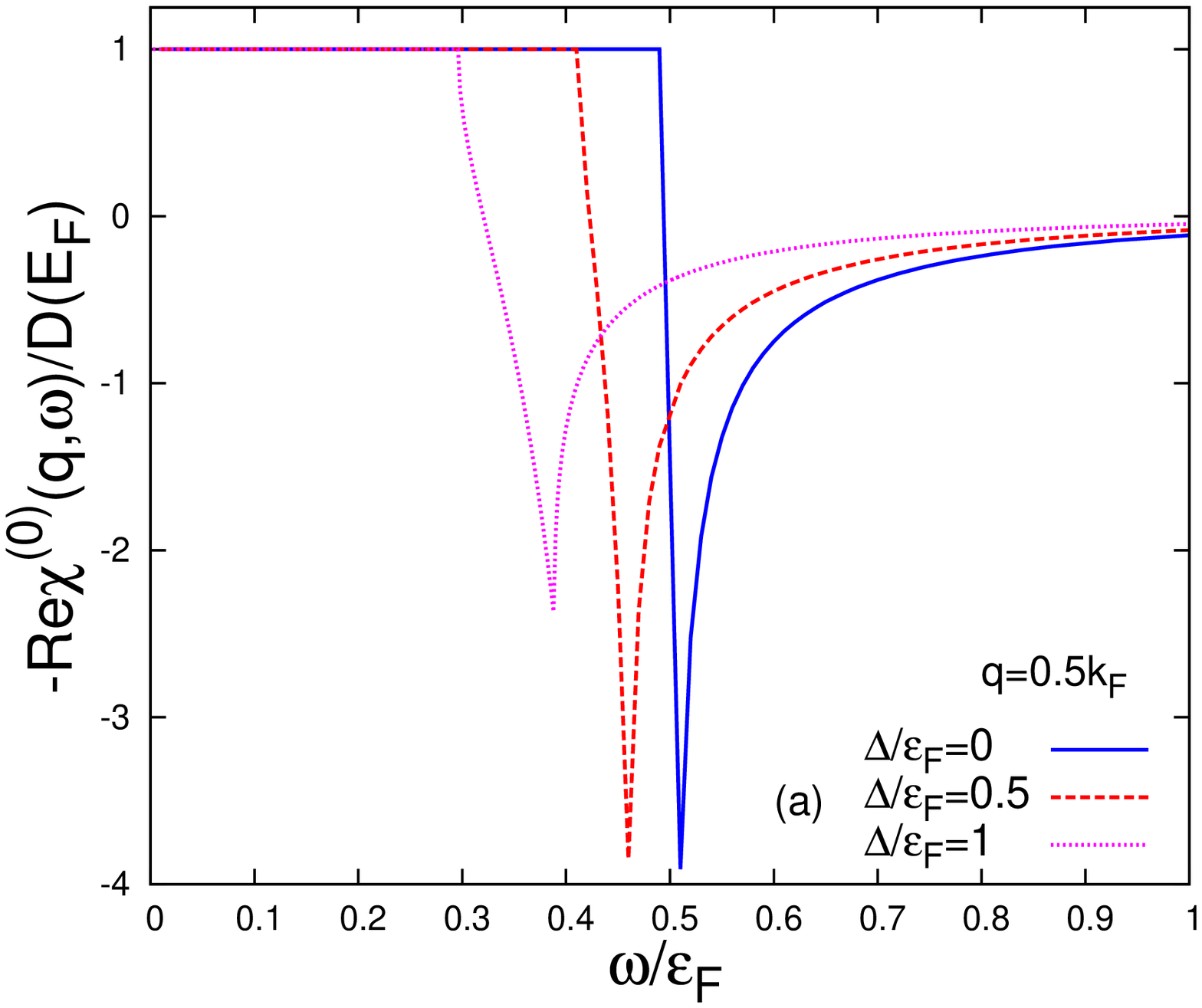}
\includegraphics[width=5cm]{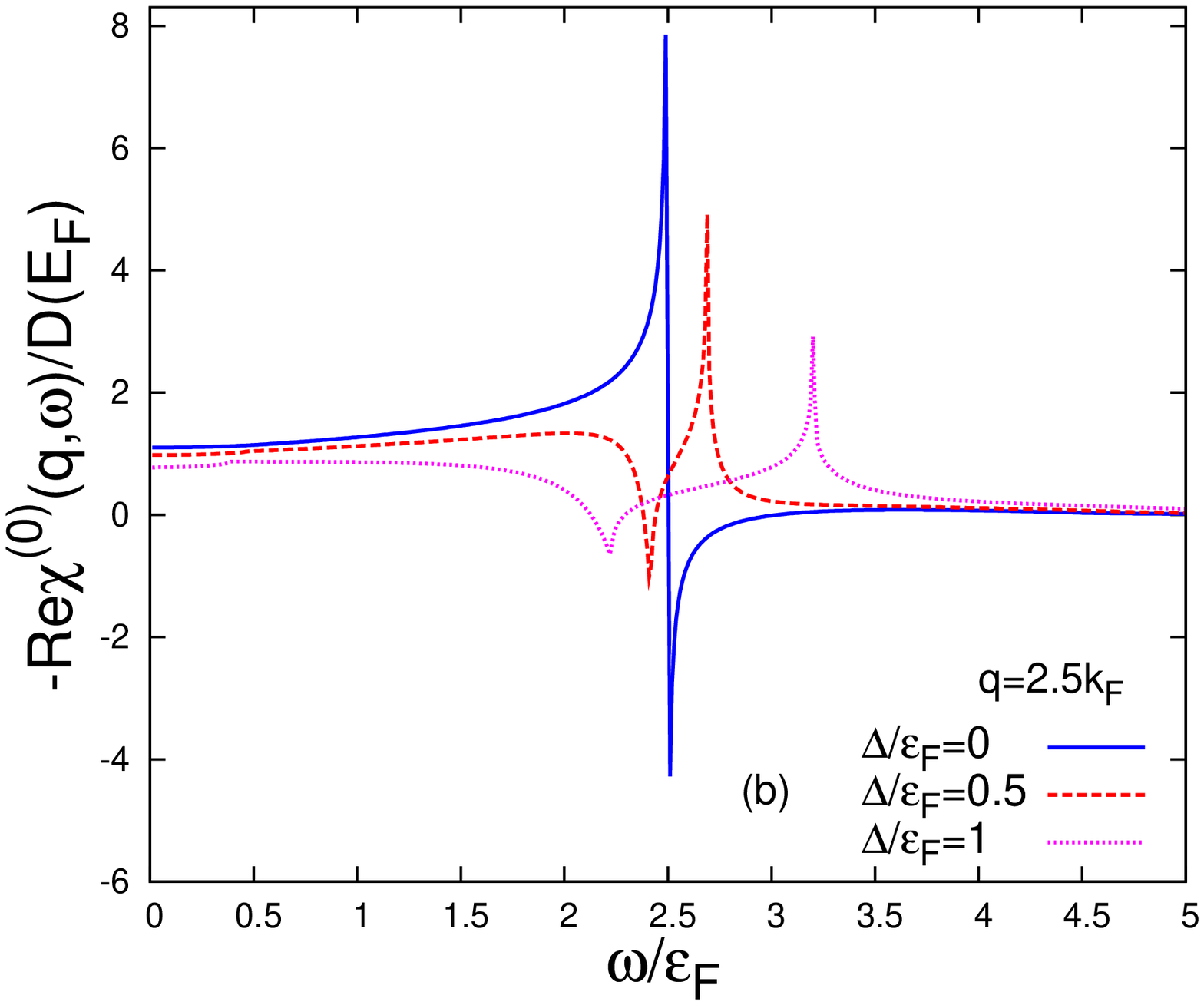}
\vfill
\includegraphics[width=5cm]{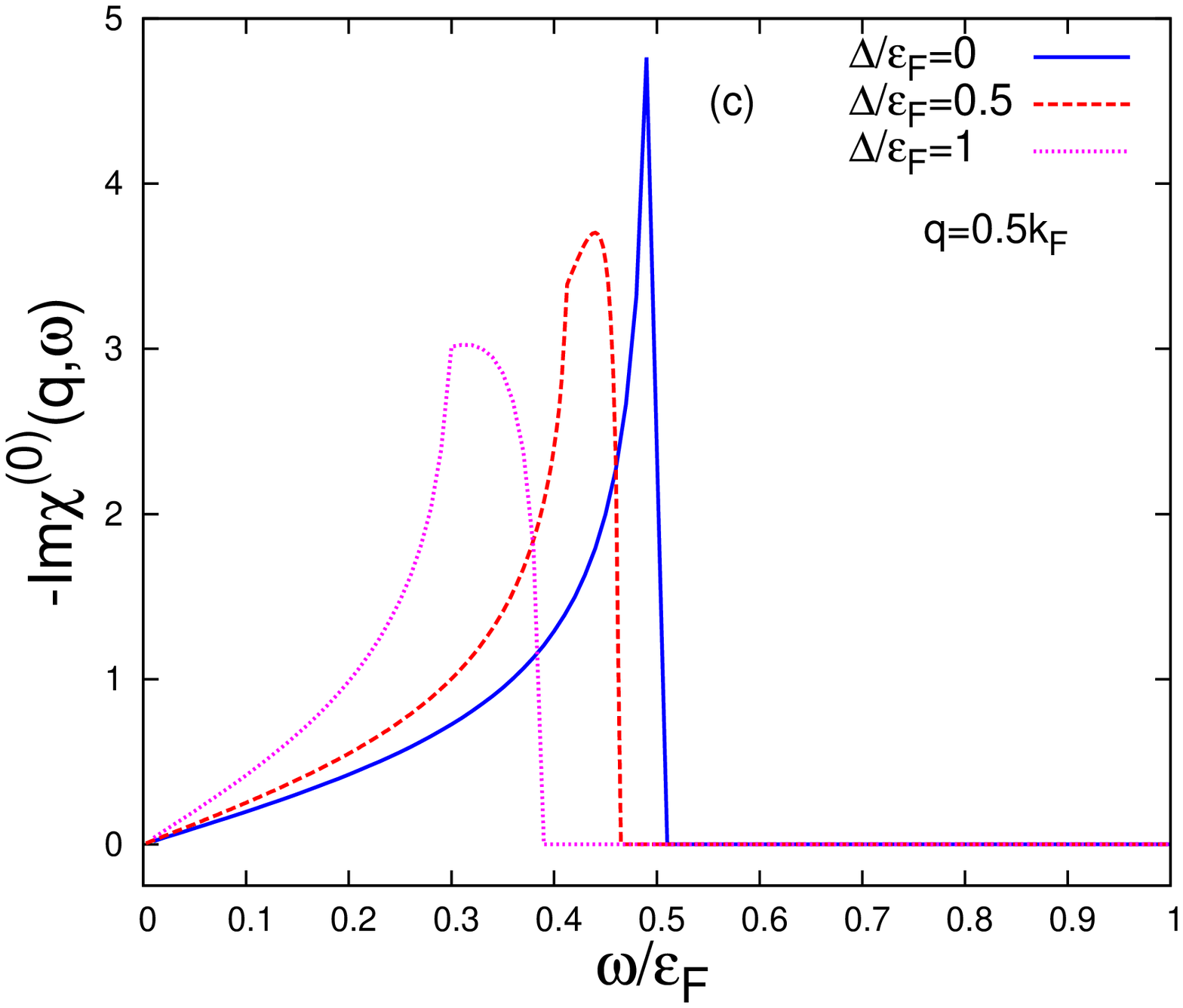}
\includegraphics[width=5cm]{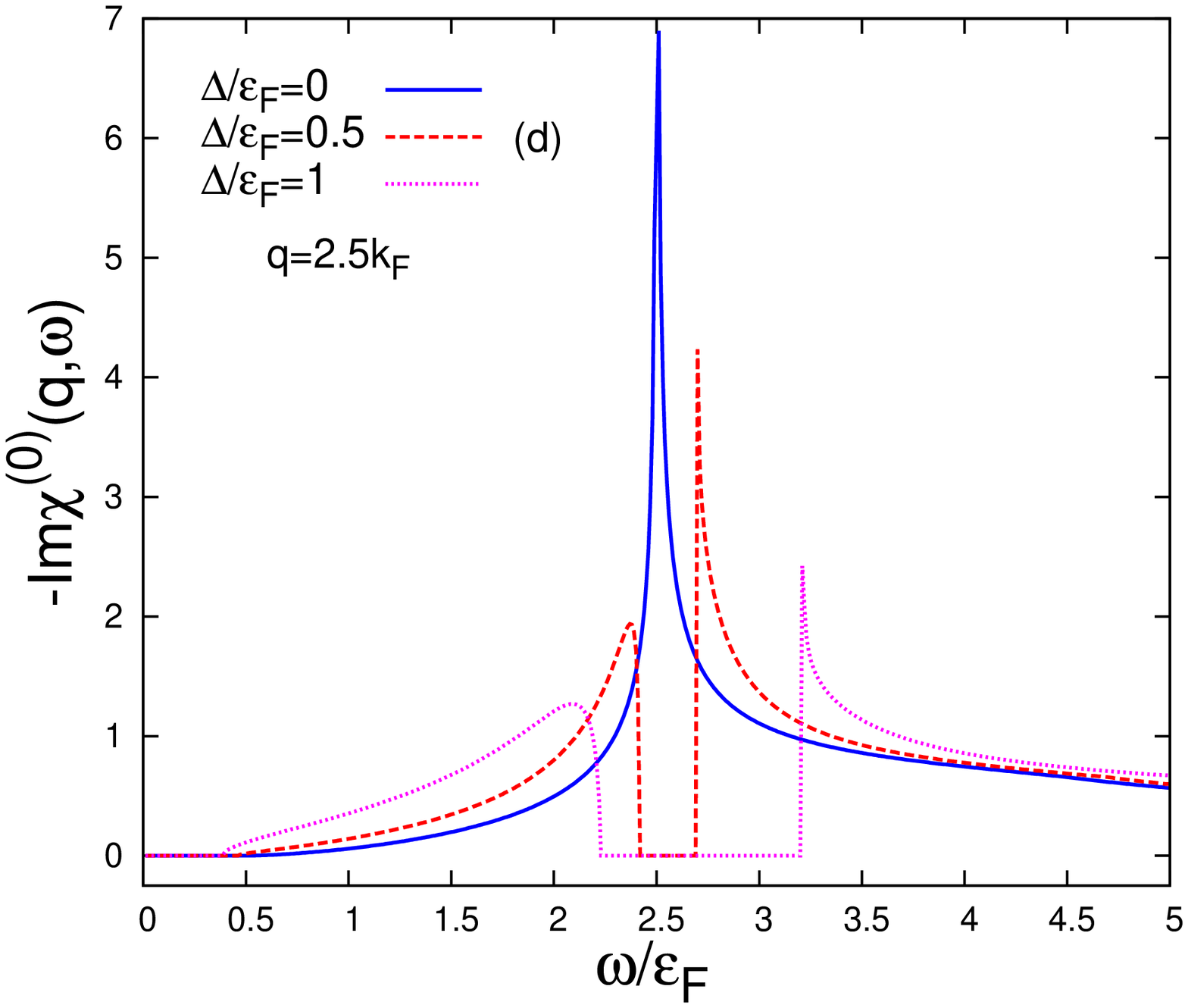}
\caption{(Color online) The gap dependence of the real
and the imaginary part of the noninteracting polarization
function as a function of $\omega$ for wavevectors (a), (c) $q=0.5 k_{\rm F}$
and (b), (d) $q=2.5k_{\rm F}$}
\end{figure}

We can calculate the total ground state energy of gapped
graphene within RPA.~\cite{alireza,yafis} The ground-state energies can be calculated using the coupling constant integration technique,
which has the contributions
$\varepsilon_{tot}=\varepsilon_{kin}+\varepsilon_{\rm xc}
$. The kinetic energy per particle is given by
$\varepsilon_{kin}=2 (E_{\rm F}^3-\Delta^3)/3\varepsilon_{\rm F}^2$.

As discussed previously~\cite{alireza,yafis} we might subtract the vacuum energy contribution from the total energy,
$\delta\varepsilon_{tot}=\varepsilon_{tot}(k_{\rm F})-\varepsilon_{tot}(k_{\rm F}=0)$.
Due to the number of states in the Brillouin zone must be
conserved, we do need a ultraviolet cut-off $k_c$, which is
approximated by $\pi k_c^2\simeq(2\pi)^2/A_0$, where $A_0$ is the
area of the unit cell. The dimensionless parameter $\Lambda$ is
defined as $k_c/k_{\rm F}\simeq (g_sg_vn^{-1}\sqrt{3}/9.09)^{1/2}\times 10^2$.\\
\begin{figure}[t]
\includegraphics[width=6cm]{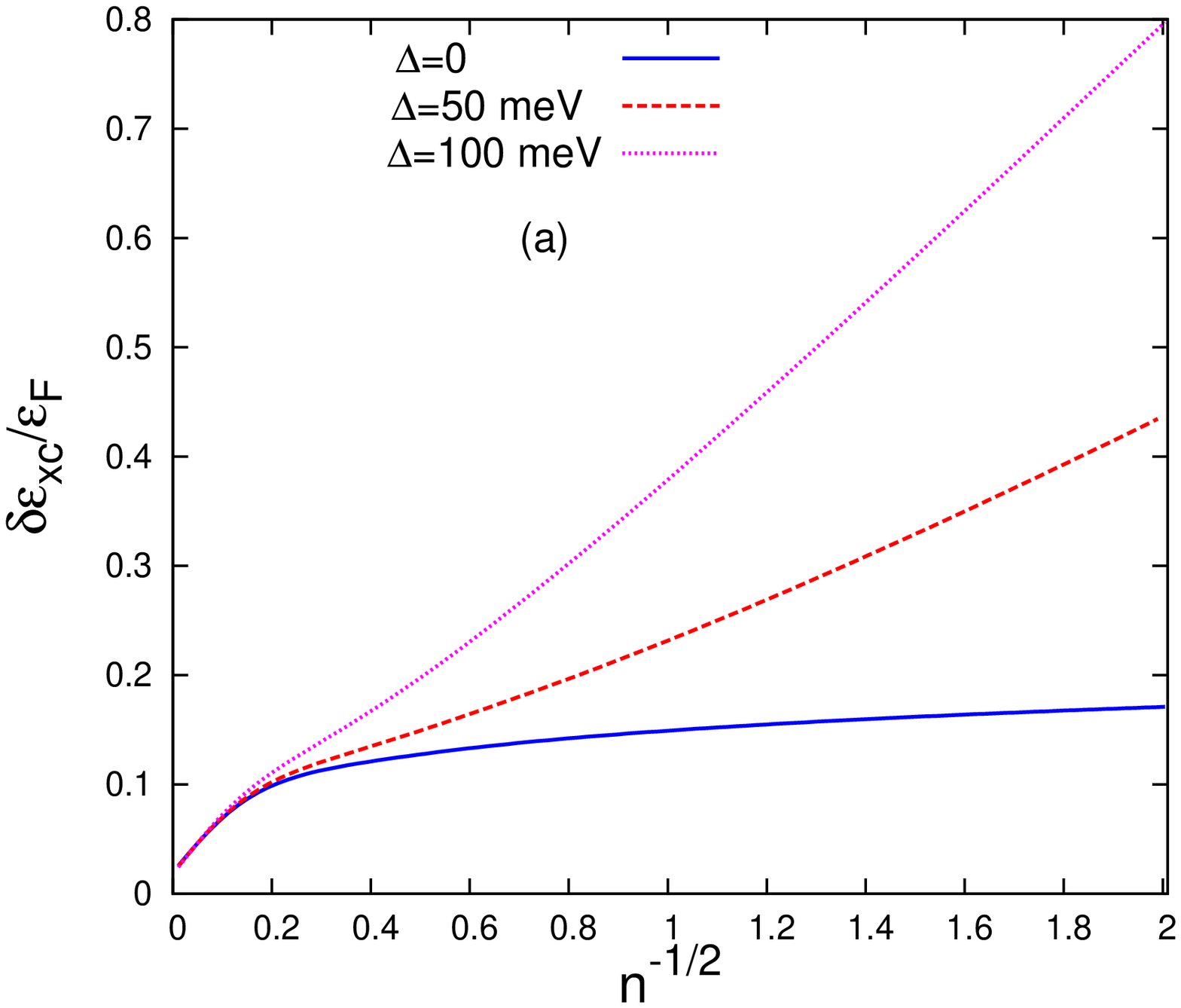}
\includegraphics[width=6cm]{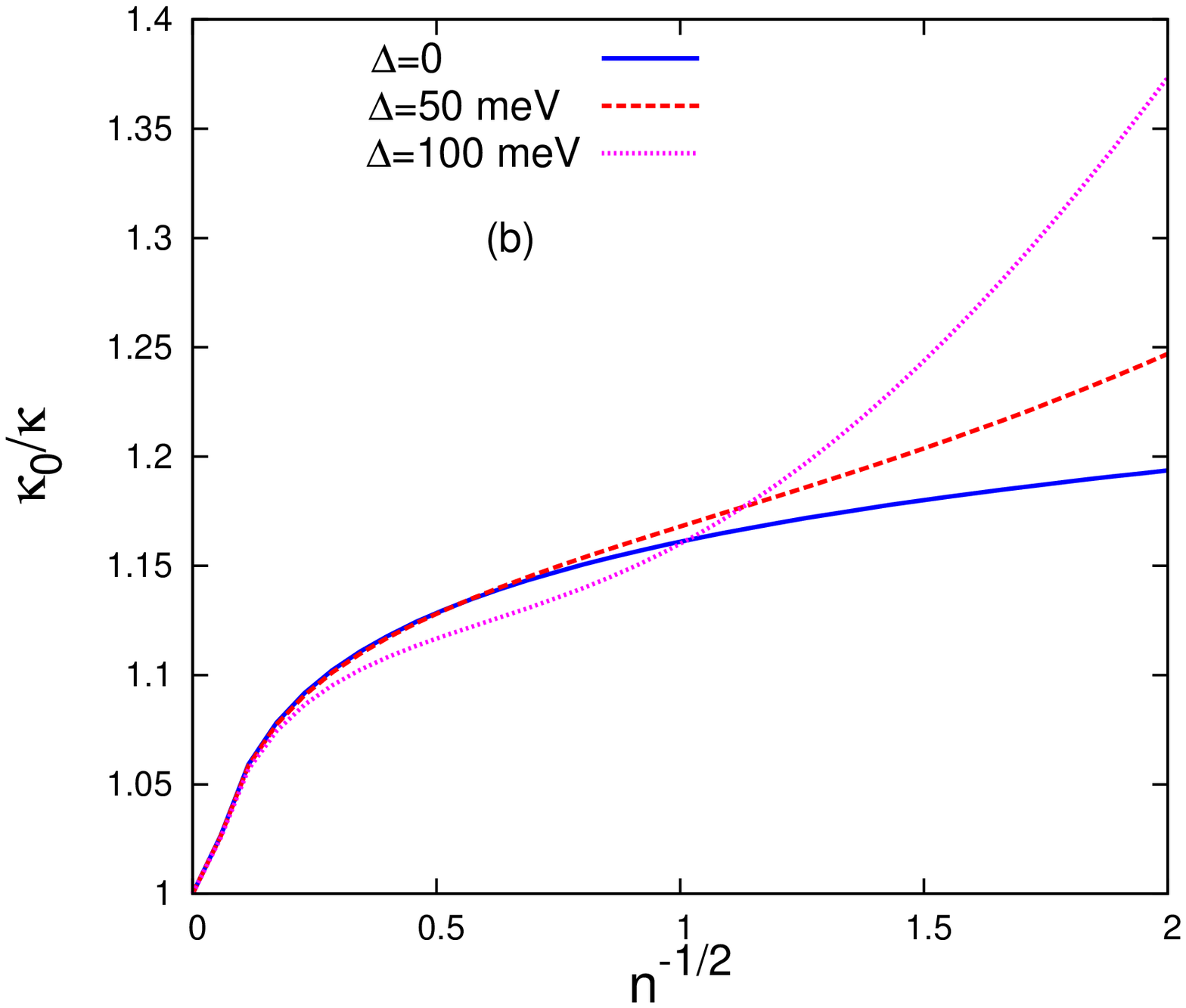}
\caption{(Color online) (a) Exchange-correlation energy and (b) compressibility
as a function of $n^{-1/2}$ ( in units of 10$^{-6}$ cm) for various $\Delta$ value.}
\end{figure}
In Fig.~3, we have shown the exchange- correlation energy in units of
$\varepsilon_{\rm F}=\hbar v_{\rm F}k_{\rm F}$, as a function of $n^{-1/2}$ in units of 10$^{-6}$ cm for
various $\Delta$ value. The exchange energy arises entirely from the antisymmetry of the many-body wave function under exchange of two electrons is positive while the correlation energy, the difference between the ground state energy and the sum of the kinetic energy and the exchange energy is negative. This has important implications on the thermodynamic properties can be calculated from the derivative of the ground state energy with respect to the density. The compressibility can be calculated from its definition, $\kappa^{-1}=n^2\partial^2(n\delta\varepsilon_{tot})/\partial n^2$.
Fig.~3(b) shows the ratio between the noninteractiong value, $\kappa_0=2/n\varepsilon_{\rm F}$ and the interaction value of compressibility as a
function of $n^{-1/2}$. The exchange tends to reduce the compressibility while correlations tends to enhance it. At large $\Delta$, a minimum structure occurs at the inverse of compressibility behavior
and we expect that at very large $\Delta$, it starts at $\kappa_0$ and reduces by increasing $n^{-1/2}$ behaves like the compressibility of the conversional 2D electron gas.

\section{THE QP SELF-ENERGY AND THE SPECTRAL FUNCTION}

The generation of QPs in an electron liquid leads to two
effects. First it induces a decay of a particle losing momentum via inelastic scattering which is determined by the
imaginary part of self-energy and second is the renormalization of
the dispersion relation of the carriers which is described by the real part of self-energy. $\Re e \Sigma^{\rm ret}({\bf k}, \omega)$ is defined
as the difference between the measured carrier energy
$\hbar\omega$, and the energy of free particle, $\xi_{s{\bf k}}=sE_{{\bf
k}}-E_{\rm F}$. To
satisfy causality, the real and imaginary parts of self-energy are
related by a Hilbert transformation. In this section, we first derive
the imaginary and the real part of QP self-energies and
then calculate some important quantities such as a renormalized Fermi velocity, a spectral function
and a band gap renormalization in the presence of
a band gap value. These quantities are related to
some important physical properties of both theoretical and
practical applications like the band structure of ARPES, the energy
dissipation rate of injected carriers and the width of the QP spectral function.~\cite{kaminski1,kaminski2}

In the G$_0$W approximation, the self-energy of gapped graphene is
given by ($\beta=1/(k_BT)$)~\cite{fateme}:
\begin{eqnarray}\label{sigma}
\Sigma_s({\bf{k}},i\omega_n)&=&-\frac{1}{\beta}\sum_{s'}\int\frac{d^2{\bf
q}}{(2\pi)^2}F^{ss'}({\bf
k,k+q})\\&\times&\sum_{m=-\infty}^{+\infty}W({\bf
q},i\Omega_m)G^{(0)}_{s'}({\bf k+q},i\omega_n+i\Omega_m)\nonumber,
\end{eqnarray}
where $W({\bf q},i\Omega_m)=V_q/\epsilon(q,i\Omega_m)$ is the
dynamical screened effective interaction and
$\epsilon(q,i\Omega_m)=1-V_q\chi^{(0)}(q,i\Omega_m)$ is dynamical
dielectric function in RPA. The overlap function for gapped
graphene $F^{ss'}({\bf k,k+q})$ arises from the graphene band
structure is given by~\cite{alireza}
\begin{eqnarray}
F^{ss'}({\bf k,k+q})= \frac{1}{2}(1+ss'\frac{\hbar^2 v_{\rm F}^2{\bf
k}\cdot({\bf k}+{\bf q})+\Delta^2}{E_{\bf k}E_{\bf k+q}}).
\end{eqnarray}
It should be noted that $F^{s=-s'}({\bf
q}=0)=0$. However, in gapless graphene, intraband backward scattering
should not be allowed, namely $F^{s=s'}({\bf q}=-2\textbf{k},\Delta=0)=0$, as
well as $F^{s=-s'}({\bf q}=0,\Delta=0)=0$. In Eq.~(\ref{sigma}),
$G_s^{(0)}({\bf k},i\omega)=1/(i\omega-\xi_{s{\bf k}}/\hbar)$ is
the noninteracting Green's function. Notice that in typical density of carriers in graphene
namely $n>10^{12}$cm$^{-2}$, the Fermi temperature is about
$T_{\rm F}=\varepsilon_{\rm F}/k_B>10^3$K, and we therefore can eliminate temperature parameter
in our calculations. To evaluate the
zero-temperature retarded self-energy we perform the line-residue
decompositions, $\Sigma^{\rm ret}_s({\bf k},\omega)=\Sigma^{\rm
line}_s({\bf k}, \omega)+\Sigma^{\rm res}_s({\bf k},\omega)$,
where $\Sigma^{\rm line}$ is obtained by performing the analytic
continuation before summing over the Matsubara frequencies, and
$\Sigma^{\rm res}$ is the correction which must be taken into
account in the total self-energy.~\cite{Giuliani} At zero temperature we have
\begin{eqnarray}
\Sigma^{\rm line}_s({\bf k}, \omega)&=&-\sum_{s'}\int\frac{d^2{\bf
q}}{(2\pi)^2}V_qF^{ss'}({\bf
k,k+q})\\&\times&\int_{-\infty}^{\infty}\frac{d\Omega}{2\pi}\frac{1}{\epsilon({\bf
q},i\Omega)}\frac{1}{\omega+i\Omega-\xi_{s'}({\bf
k+q})/\hbar},\nonumber
\end{eqnarray}
and
\begin{eqnarray}\label{si1}
\Sigma^{\rm res}_s({\bf k}, \omega)&=&\sum_{s'}\int\frac{d^2{\bf
q}}{(2\pi)^2}\frac{V_q}{\epsilon({\bf q},\omega-\xi_{s'}({\bf
k+q})/\hbar)}F^{ss'}({\bf
k,k+q})\nonumber\\&\times&[\Theta(\omega-\xi_{s'}({\bf
k+q})/\hbar)-\Theta(-\xi_{s'}({\bf k+q})/\hbar)].
\end{eqnarray}
The line contribution of the self-energy is
purely real. The imaginary part of the self-energy has two contributions where $\Im m\Sigma_+^{\rm
ret}({\bf{k}},\omega)=\Im m\Sigma_{+,\rm intra}^{\rm
res}({\bf{k}},\omega)+\Im m\Sigma_{+,\rm inter}^{\rm
res}({\bf{k}},\omega)$, and real part of the self energy can be decomposed
as $\Re e\Sigma_+^{\rm ret}({\bf{k}},\omega)=\Sigma_{+}^{\rm
line}({\bf{k}},\omega)+\Re
e\Sigma_{+,\rm inter}^{\rm res}({\bf{k}},\omega)+\Re e\Sigma_{+,\rm intra}^{\rm res}({\bf{k}},\omega)$.

For $\omega > 0$ and fixed ${\bf q}$, the RPA decay process represents scattering of an
electron from momentum ${\bm k}$ and energy $\omega$ to ${\bm k}+{\bm q}$ and $\xi_{s'}({\bm k}+{\bm q})$, with all
energies in Eq.~(\ref{si1}) measured from the Fermi energy of doped graphene. Since the Pauli exclusion principle requires that the final state
is unoccupied, it must lie in the conduction band, {\it i.e.} $s'=+1$. Furthermore since the Fermi sea is initially in its
ground state, the QP must lower its energy, {\em i.e.} $\xi_{s'} < \omega$, electrons decay by going down in energy.
For $\omega < 0$, the self-energy expresses the decay of holes inside the Fermi sea, which scatter to a final state, by exciting
the Fermi sea. In this case the final state must be occupied so both band indices are allowed for $s'$, and
energy conservation requires that holes decay by moving up in energy. Since photoemission measures the properties of
holes produced in the Fermi sea by photo ejection, only $\omega < 0$ is relevant for this experimental probe.

In what follows, we calculate the intraband and interband contributions of self-energy. We have found the intraband term
of residue part of self-energy as following for various values
of the frequencies,
\begin{eqnarray}
&&\Sigma_{+,\rm intra}^{\rm res}({\bf
k},\omega>0)= C \int_{\rm
max(0,~k_{\rm F}-k,~k-\beta)}
^{k+\beta} dq\int^{\rm
min(\hbar\omega+E_{\rm F},~\alpha_{+})} _{\rm
max(E_{\rm F},~\alpha_{-})}dy f_+(y,q)\nonumber\\
&&\Sigma_{+,~\rm intra}^{\rm res}({\bf
k},\Delta-E_{\rm F}<\hbar\omega<0)=-C
\int_{\rm max(0,~k-k_{\rm F},~\beta-k)}^{k+k_{\rm F}} dq\int^{\rm
min(E_{\rm F},~\alpha_{+})}_{\rm max(0,~\hbar\omega+E_{\rm F}
,~\alpha_{-})}dy f_+(y,q)\nonumber\\
&&\Sigma_{+,\rm intra}^{\rm res}({\bf
k},\hbar\omega<-(\Delta+E_{\rm F}))=-C
\int_{\rm max(0,~k-k_{\rm F})}^{k+k_{\rm F}} dq\int^{\rm
min(E_{\rm F},~\alpha_{+})}_{\rm max(0,~\hbar\omega+E_{\rm F}
,~\alpha_{-})}dy f_+(y,q)~,
\end{eqnarray}
where
\begin{equation}
f_{\pm}(y,q)=\frac{\pm(y\pm E_k)^2\mp
q^2}{\epsilon({\bf q},\omega+E_{\rm F}\mp
y)\sqrt{4k^2q^2-(y^2-E_k^2-q^2)^2}}~\nonumber,
\end{equation}
$C=e^2/2\pi\epsilon E_k$, $\alpha_{\pm}=\sqrt{\hbar^2 v^2_{\rm F}(k \pm q)^2+\Delta^2}$ and $\beta=\sqrt{\hbar^2 \omega^2+\hbar^2 v^2_{\rm}k^2_{\rm F}+2\hbar\omega E_{\rm F}}$.
On the other hand, the interband contribution of residue part of the self energy is determined by
\begin{equation}
\Sigma_{+,\rm inter}^{\rm res}({\bf
k},\hbar\omega<-(\Delta+E_{\rm F}))=-C
\int_{\rm max(0,~k-\beta)}^{k+\beta}
dq\int_{\alpha_{-}}^{\rm
min(\alpha_{+},-\hbar\omega-E_{\rm F})}dy f_{-}(y,q)~,
\end{equation}
and eventually for the line contribution of self-energy we have

\begin{eqnarray}
\Sigma_{+,\rm intra}^{\rm
line}({\bf{k}},\omega)&=&-\frac{e^2}{4\pi^2\epsilon}\int_0^{k_c}
dq \int_{0}^{2\pi}d\phi F^{++}({\bf
q,q+k},\Delta)\int_{-\infty}^{+\infty}d\Omega
\frac{g_{+}(\phi,\Omega,q)}{\epsilon({\bf q},i\Omega)}\nonumber\\
\Sigma_{+,\rm inter}^{\rm
line}({\bf{k}},\omega)&=&-\frac{e^2}{4\pi^2\epsilon}\int_0^{k_c}
dq \int_{0}^{2\pi}d\phi F^{+-}({\bf
q,q+k},\Delta)\int_{-\infty}^{+\infty}d\Omega
\frac{g_{-}(\phi,\Omega,q)}{\epsilon({\bf q},i\Omega)}
\end{eqnarray}
where
\begin{equation}
g_{\pm}(\phi,\Omega,q)=\frac{\hbar\omega+E_{\rm F}\mp E_{\bf
{k+q}}}{(\hbar\omega+E_{\rm F}\mp E_{{\bf k+q}})^2+\hbar^2\Omega^2}
\end{equation}
and $\phi$ denotes an angle between ${\bf k}$ and ${\bf q}$.
Note that the real part of self-energy is
$k_c$ dependent.

Now we are in a situation that can calculate some important
physical quantities.
The QP lifetime or the single-particle relaxation time $\tau$, is
obtained by setting the frequency to the on-shell energy in imaginary part of the
self-energy, $\tau_s^{-1}=\Gamma_s({\bf k},{\xi_{s\bf
k}}/\hbar)=\frac{2}{\hbar}|\Im m \Sigma_s^{\rm ret}({\bf
k},\xi_{s\bf k}/\hbar)|$
where $\Gamma_s({\bf k},{\xi_{s\bf k}}/\hbar)$ is the quantum
level broadening of the momentum eigenstate $|s{\bf k}>$. This
quantity is identical with the Fermi's golden rule expression for
the sum of the scattering rate of a QP and quasihole at
wavevector ${\bf k}$.~\cite{Giuliani} From Eqs. 8 and 9, one can conclude that
total contribution of the imaginary part of the retarded
self-energy on the energy shell comes from the intraband term,
$\Im m \Sigma_+^{\rm ret}({\bf
k},\xi_{\bf k}/\hbar)=\Im m\Sigma^{\rm res}_{\rm intra}({\bf k}, \xi_{\bf k}/\hbar)$.~\cite{fateme}
In the case of gapless graphene, scattering rate is a
smooth function because of the absence of both plasmon emission
and interband processes.~\cite{inelastic1,inelastic2} However, with generating a
gap and increasing the amount of it, plasmon emission cause
discontinuities in the scattering time, similar to conventional
2D electron gas.~\cite{asgari1,asgari2} We have thus two mechanisms for scattering of
the QPs. The excitation of electron-hole pairs which is
dominant process at long wavelength regions and the excitation of plasmon
appears in a specific wave vector. As discussed
previously~\cite{fateme}, in clean graphene sheets the inelastic mean free path reduces by increasing
the gap whereas the mean free path is large enough in the range
of the typical gap values 10-130 meV, and thus transport remains in the semi-ballistic regime.

The many-body interactions in graphene as a function of doping can be observed by ARPES which plays as a central role to investigate QP properties such as group velocity and lifetime of carriers on the Fermi surface. ARPES is a useful complementary tool which capable of measuring the constant energy surfaces for all partially occupied states and the fully occupied band structure. The information of band dispersion and
the Fermi surface can be elicited from those data measured in ARPES experiments. The relation of the Green's function to the single-particle excitation spectrum in the interacting fluid is expressed by its spectral function. The spectral function is related to the retarded self-energy by the following expression~\cite{Giuliani}
\begin{eqnarray}
A_s({\bf k},\omega)=\frac{\hbar}{\pi}\frac{|\Im m\Sigma_s^{\rm
ret}({\bf k},\omega)|}{[\hbar\omega-\xi_s({\bf k})-\Re
e\delta\Sigma_s^{\rm ret}({\bf k,\omega})]^2+[\Im m\Sigma_s^{\rm
ret}({\bf k},\omega)]^2}
\end{eqnarray}
where $\delta \Sigma_s^{\rm ret}({\bf k},\omega)=\Sigma_s^{\rm
ret}({\bf k},\omega)-\Sigma_s^{\rm ret}(k_{\rm F},0)$, and then ARPES
intensity can be described by $I({\bf k},\omega)=A({\bf
k},\omega)n(\omega)$, where $n(\omega)$ is the Fermi-Dirac
distribution. The spectral function is the Lorentzian function
where $\Re e\Sigma$ specifying the location of the peak of the
distribution, and $|\Im m\Sigma|$ is the linewidth. The amplitude
of the the Lorentzian function is proportional to $1/|\Im
m\Sigma|$. This quantity is the distribution of energies
$\hbar\omega$, in the system when a QP with momentum
${\bf k}$, is added or removed from that. For the noninteracting
system we get $A^{(0)}({\bf k},\omega)=\delta(\omega-\xi({\bf
k})/\hbar)$. The Fermi liquid theory applies only when the
spectral function at the Fermi momentum $A^{(0)}(k=k_{\rm F},\omega)$,
behaves as a delta function, and has a broadened peak
indicating damped QPs at $k\neq k_{\rm F}$.


%

To progress of the interband single particle excitation and
plasmon effects on the $\Im m \Sigma_s^{\rm ret}$, we must study the
retarded self-energy on the off-shell frequency which is $\omega\neq\xi_{s{\bf
k}}/\hbar$.~\cite{im1,im2} This quantity gives the scattering rate of a
QP with momentum ${\bf k}$ and kinetic energy
$\hbar\omega+E_{\rm F}$. The scattering rate or the linewidth raising
from electron-electron interactions is anisotropic and varies
significantly via wavevector at a constant energy. The imaginary
part of self energy shows the width of the QP spectral
function.

In Fig.~4 we have shown the absolute value of the imaginary
part of the self energy in unit of $\varepsilon_{\rm F}$ for various gap values. It would be noticed that there is an area of frequency is associated to the gap value, $2\Delta$ in which no QP could enter in. In this case, there is a gap in the $\Im m \Sigma$ between $\xi_{-,k=0}$ and $\xi_{+,k=0}$. We see that $\Im m \Sigma_{+}$ vanishes as $\omega^2$ for $\omega$ tends to zero, a universal properties of normal Fermi liquid. Moreover, at large frequency, $\Im m \Sigma_{+}$ tend towards to $\omega$ linearly. Except from the Dirac point, the conduction band $\Im m \Sigma_{+}$ peaks broaden because of the dependence on scattering angle of $\xi({\bf k}+{\bf q})$. For low energy, only intraband single particle excitation contributes to $\Im m \Sigma$ up to $E_{\rm F}$ and then the interband single particle excitation contribution increases sharply about $E_{\rm F}$. The interband contribution increases with increasing the gap values.

To evaluate the scattering rate in interband channel, we have shown $\Im m \Sigma_{inter(intra)}$ as a function of frequency in Fig.~5. The intraband contribution of the
imaginary part of self energy associated to scattering rate of QP in the intraband contribution increases with increasing the gap values while the interband
contribution reduces, as we physically expected. Moreover, by increasing of the
electrons in the conduction band the interband scattering
rate reduces whereas the intraband scattering contribution increases. The gap value suppresses the scattering rate at $\omega=-\varepsilon_{\rm F}$.

In Fig.~6 we have plotted the
real part of self-energy in unit of $\varepsilon_{\rm F}$ as a
function of the energy for various gap values. Notice again that the real part of residue self-energy has a gap which is associated
the feature calculated in the imaginary part of self-energy. The line part of self energy is a continues curve and then we have a jump near to the boundary of gap values in the $\Re e \Sigma$ for gapped graphene. A kink around $E_{\rm F}$ is associated to the interband plasmon contribution and it is broaden due to the gap value. This feature affects noticeably in the interacting electron density of states.

\begin{figure}[h]
\includegraphics[width=6cm]{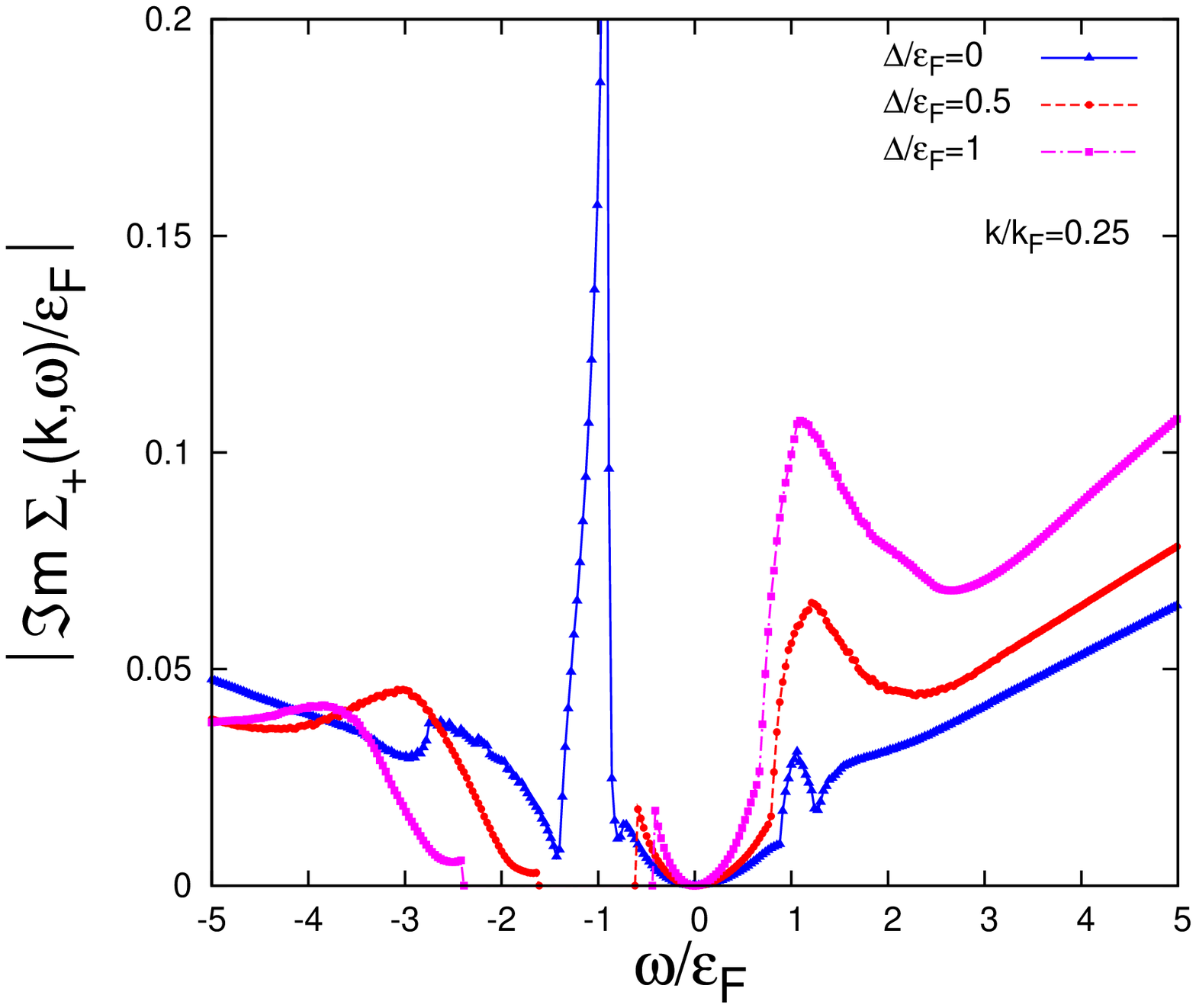}
\includegraphics[width=6cm]{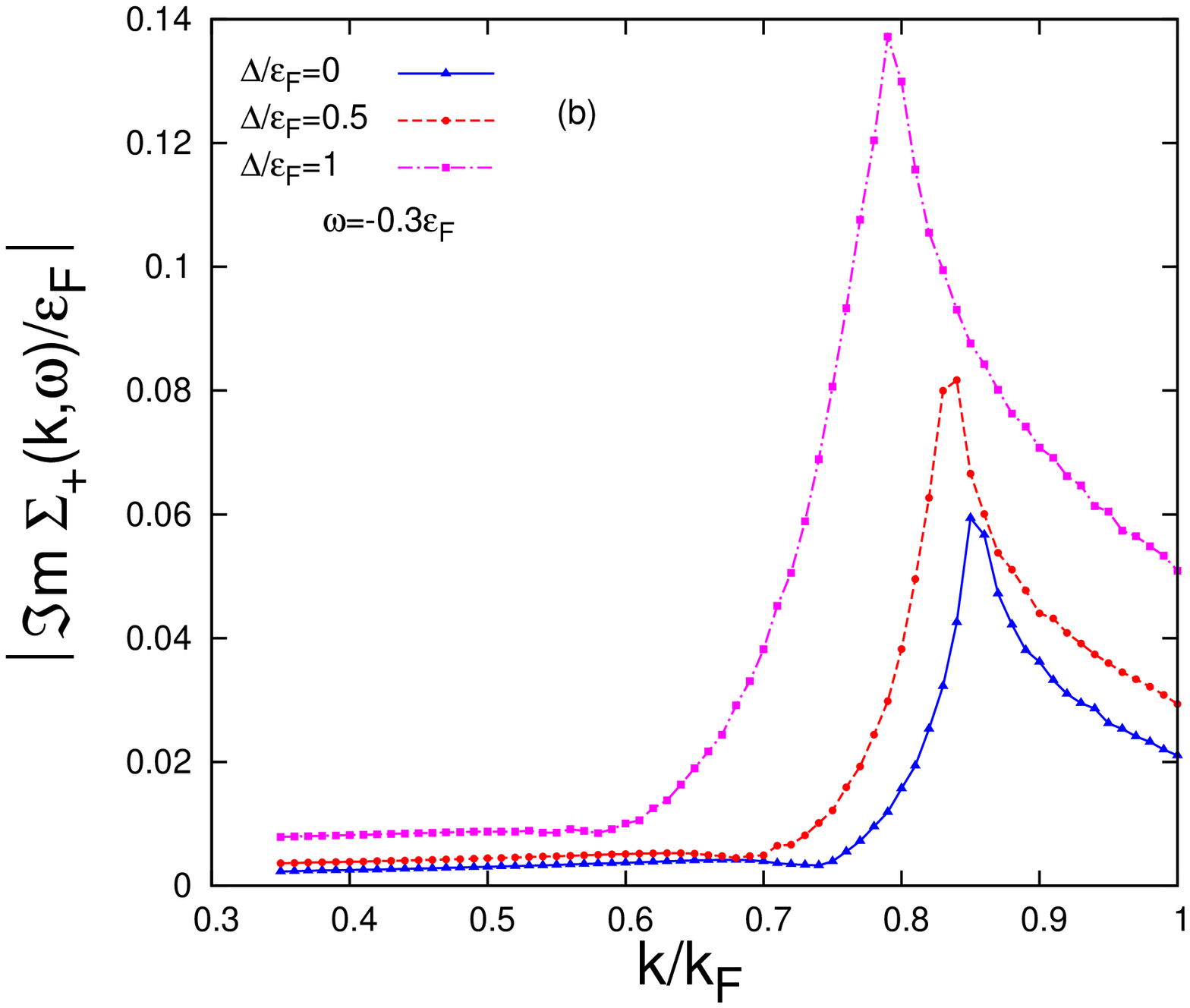}
\caption{(Color online) The absolute value of the imaginary part
of retarded self-energy ( $+$ channel) as a function of (a) $\rm \omega$ and (b)
$\rm k$, for the various energy gaps.} \label{}
\end{figure}
\begin{figure}[h]
\includegraphics[width=6cm]{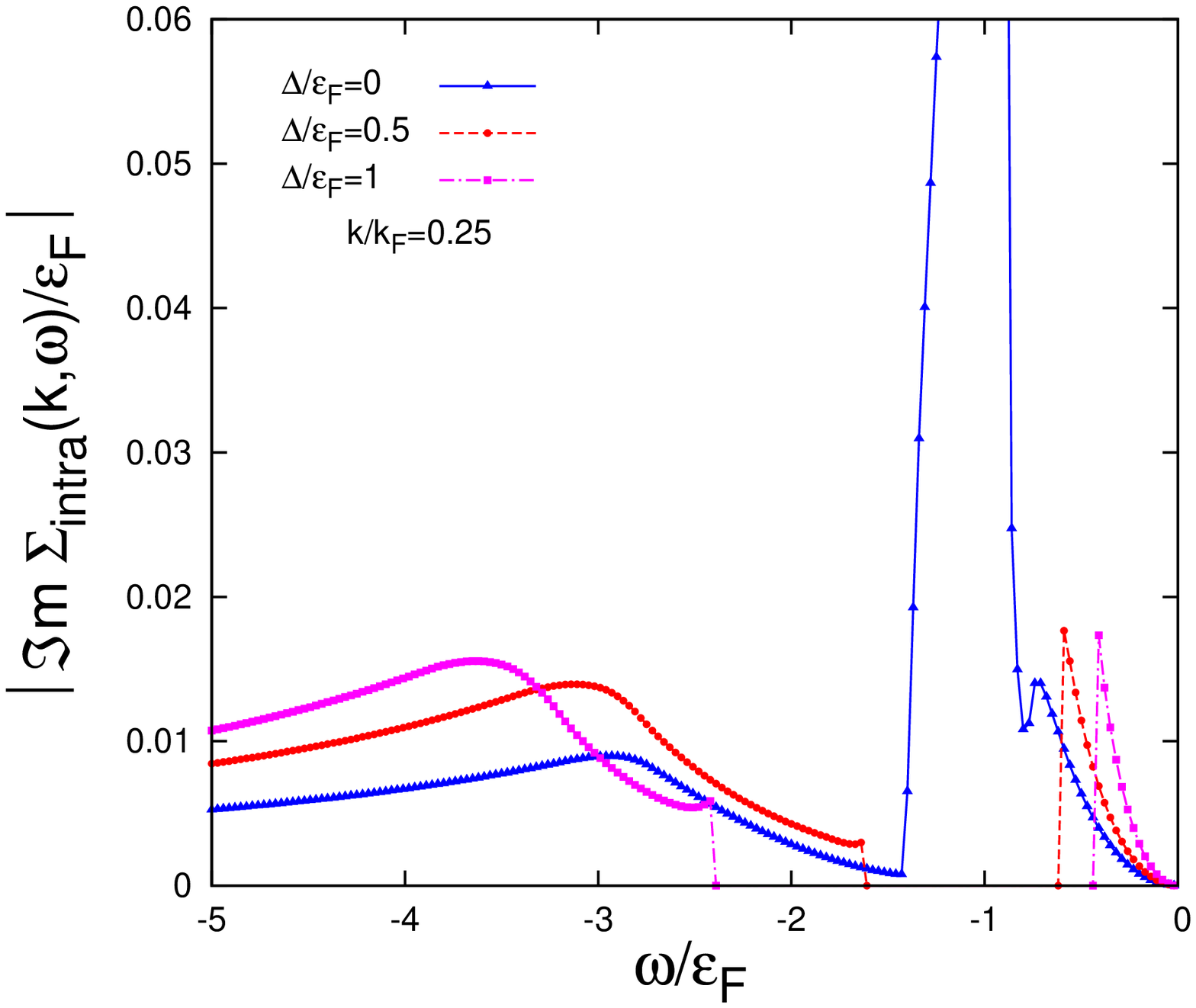}
\includegraphics[width=6cm]{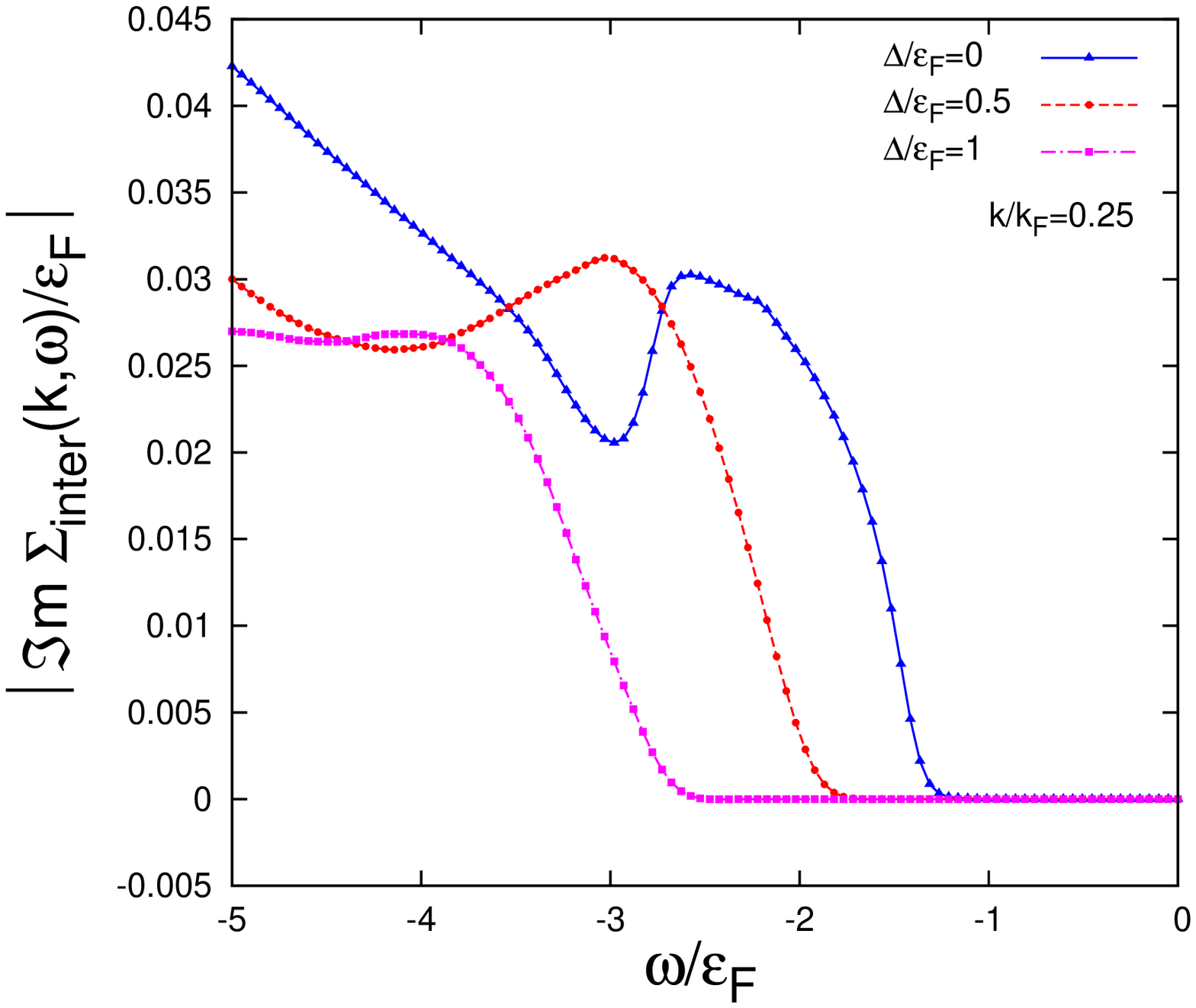}
\caption{(Color online) (a) The intraband and (b) the interband
contributions of the imaginary part of self-energy ( $+$ channel) as a
function of $\omega$ for the various energy gaps at $\rm
k=0.25k_{\rm F}$.}
\end{figure}
\begin{figure}[h]
\includegraphics[width=6cm]{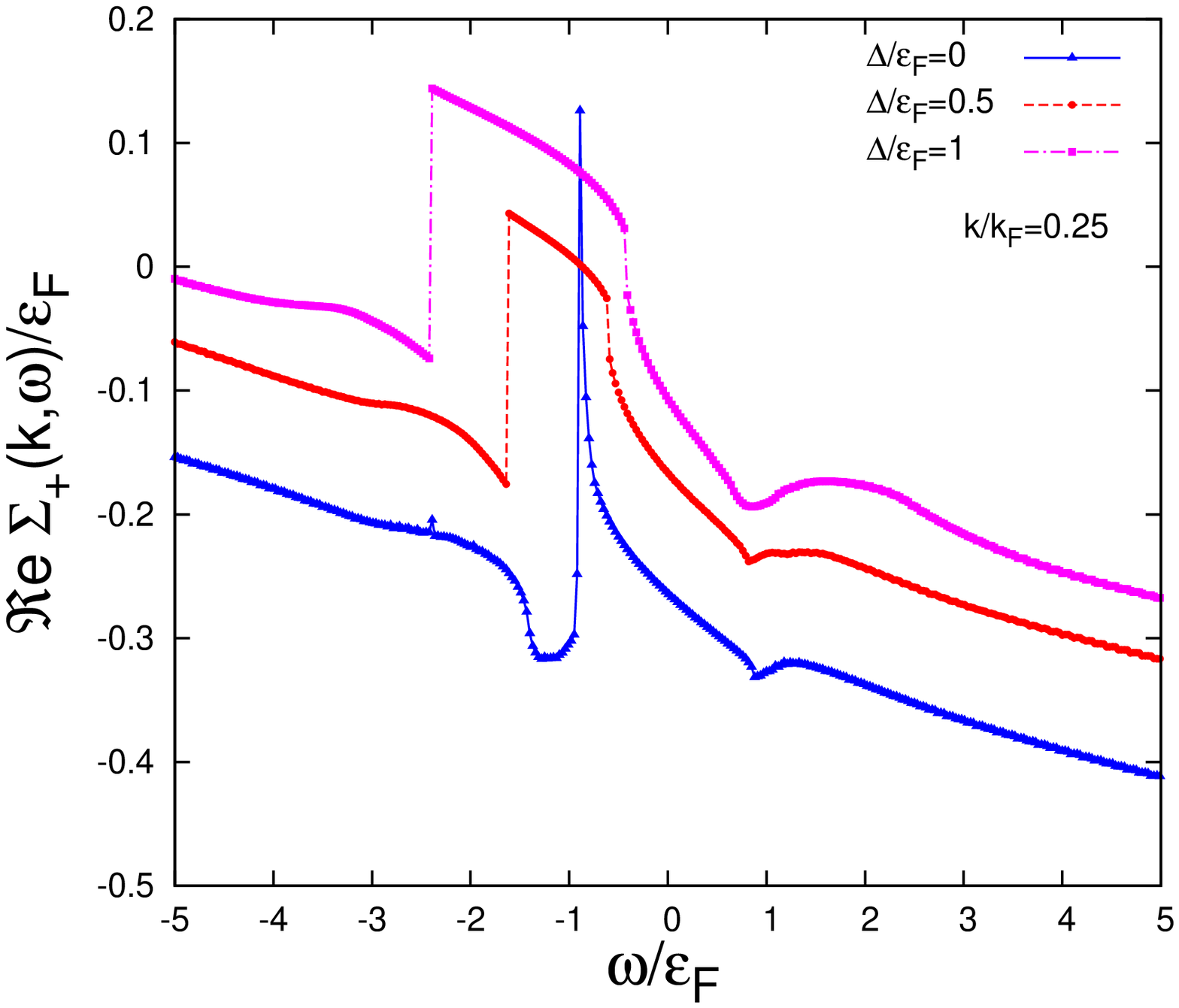}
\includegraphics[width=6cm]{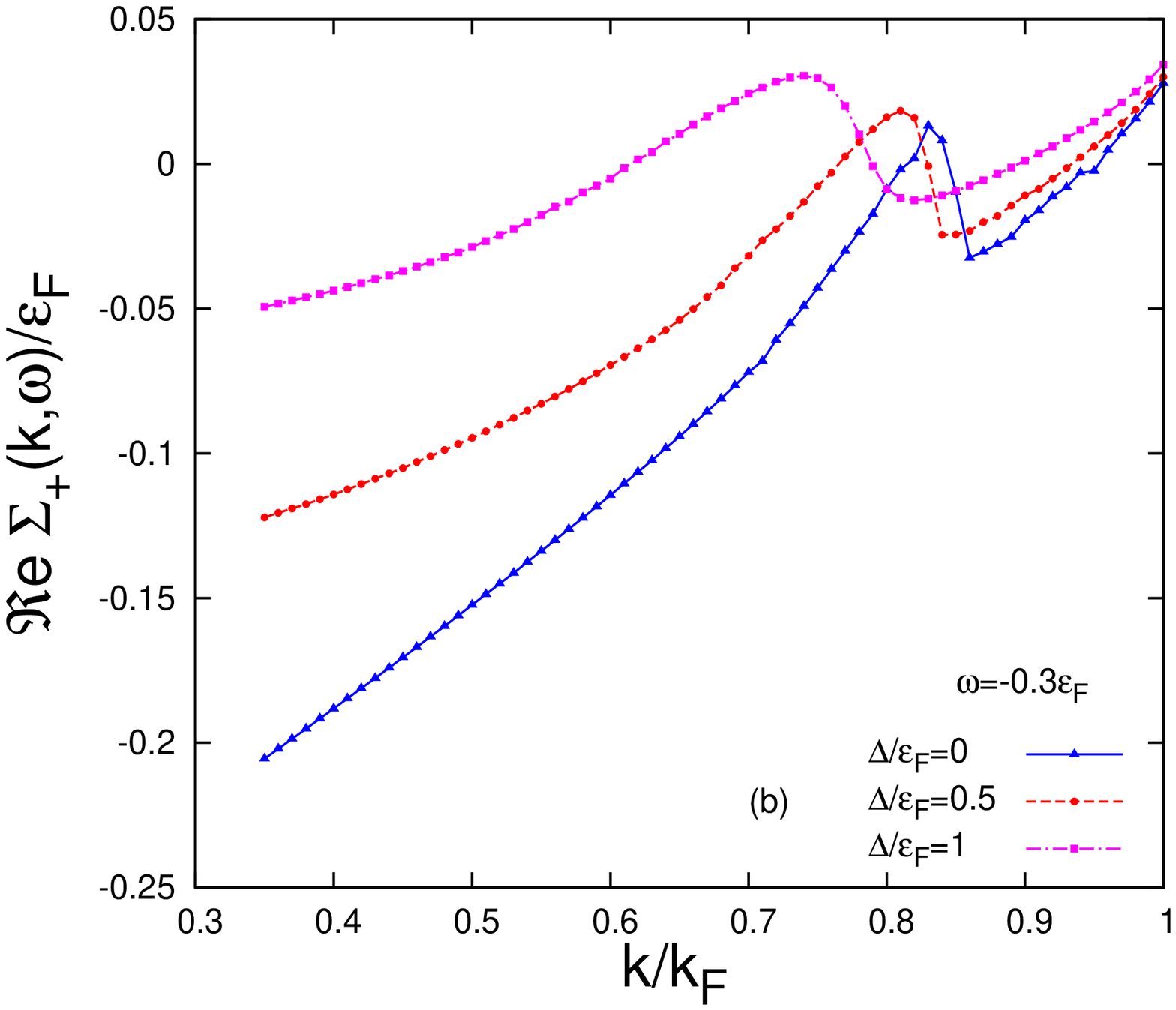}
\caption{(Color online) The real part of retarded self-energy ( $+$ channel)
as a function of (a) $\omega$ and (b) $\rm k$, for the various
energy gaps at $\Lambda=100$. The $\Re e \Sigma_+$ are measured
from the interaction contribution of the chemical potential $\Re e
\Sigma_+(k_{\rm F},\omega=0)$.} \label{}
\end{figure}
\begin{figure}[h]
\includegraphics[width=6cm]{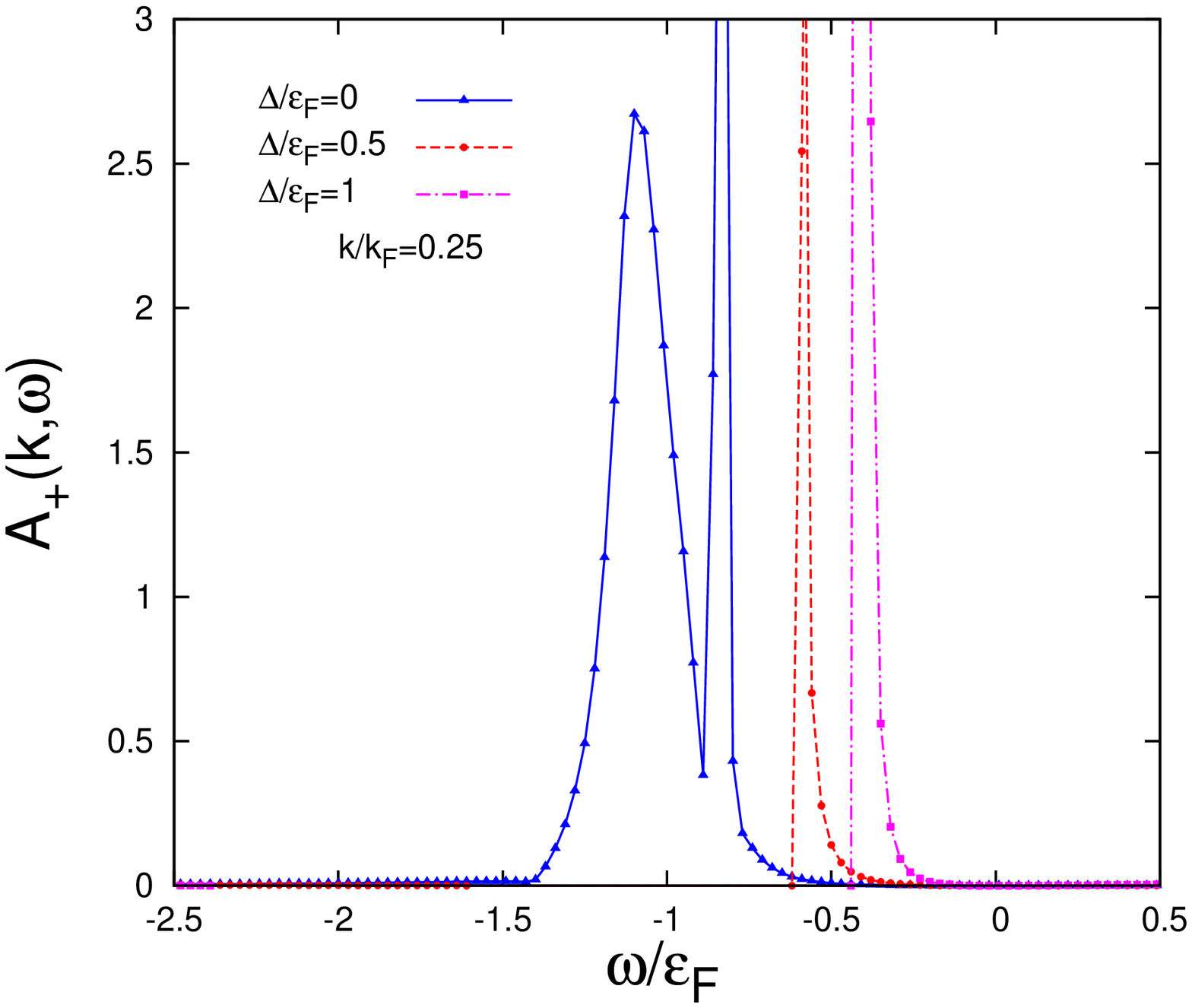}
\includegraphics[width=6cm]{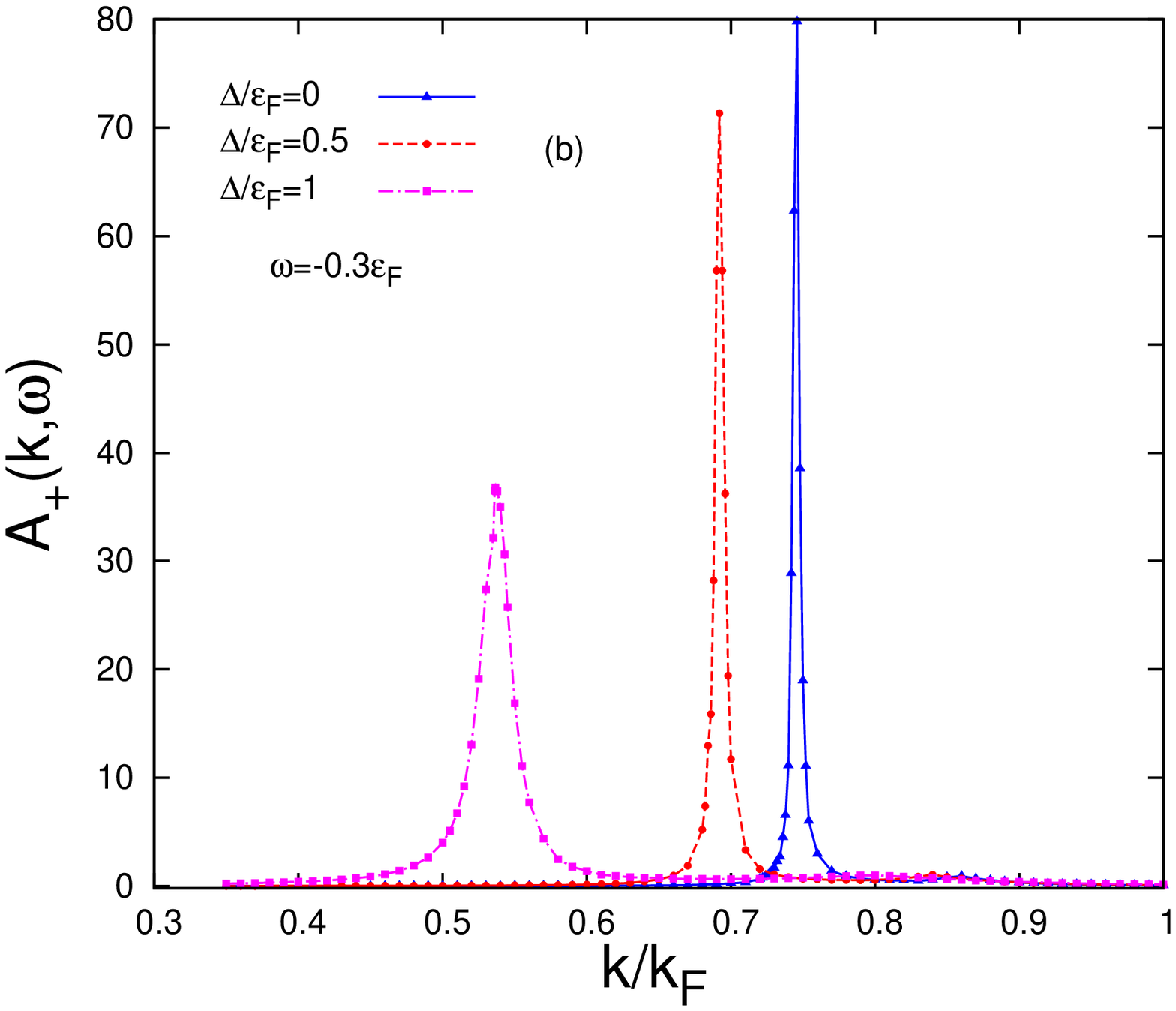}
\caption{(Color online) The QP spectral function for $+$ channel as a
function of $\omega$ (a) and $\rm k$ (b), for the various energy
gaps at $\Lambda=100$.} \label{}
\end{figure}

As discussed before~\cite{im1,im2} in a zero
temperature and disorder free gapless graphene, the peaks of the spectral function correspond to the nearly solutions of Dyson's equation in which the quasiparticle excitation energies are obtained by
$E=\xi_{+}+\Re
e\delta\Sigma_+^{\rm ret}$. The intersection of $\Re e \Sigma$ and the lines $E-\xi_+$ indicates a satellite long wavelength plasmaron peak related to the
electron-plasmon excitation due to the long-range electron-electron Coulomb interaction and the Dyson equation with $\Im m\Sigma=0$ corresponds to a QP peak related to the single particle excitation. Importantly, in the presence of gap values, the plasmaron peak suppressed. In Figs.~7(a) and (b) we have shown the
energy distribution curves (EDC) and momentum distribution curves
(MDC), respectively. In the presence of gap values, as shown in
Fig.~7(a) there is only the single QP peak.
\begin{figure}[h]
\includegraphics[width=5cm]{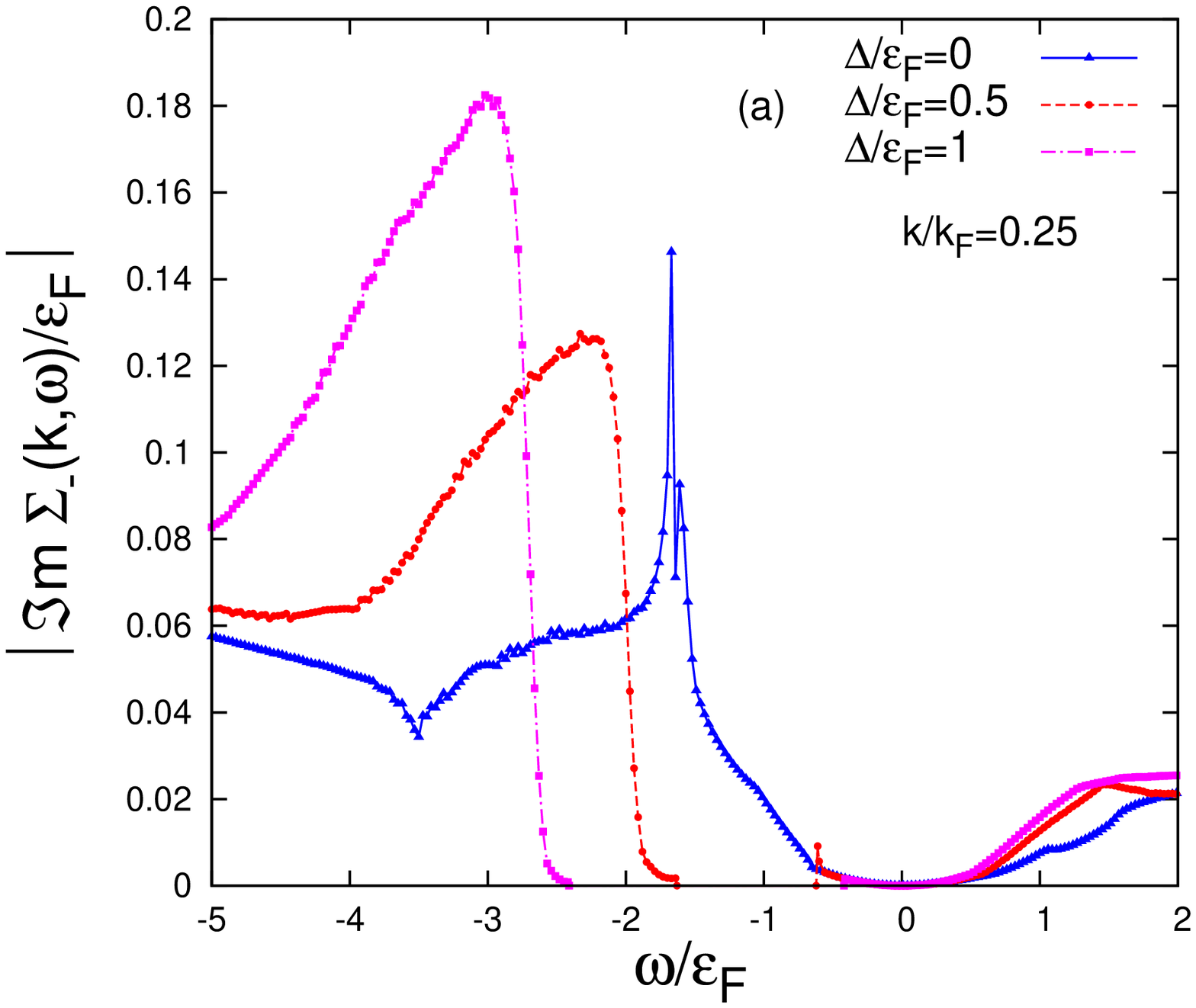}
\includegraphics[width=5cm]{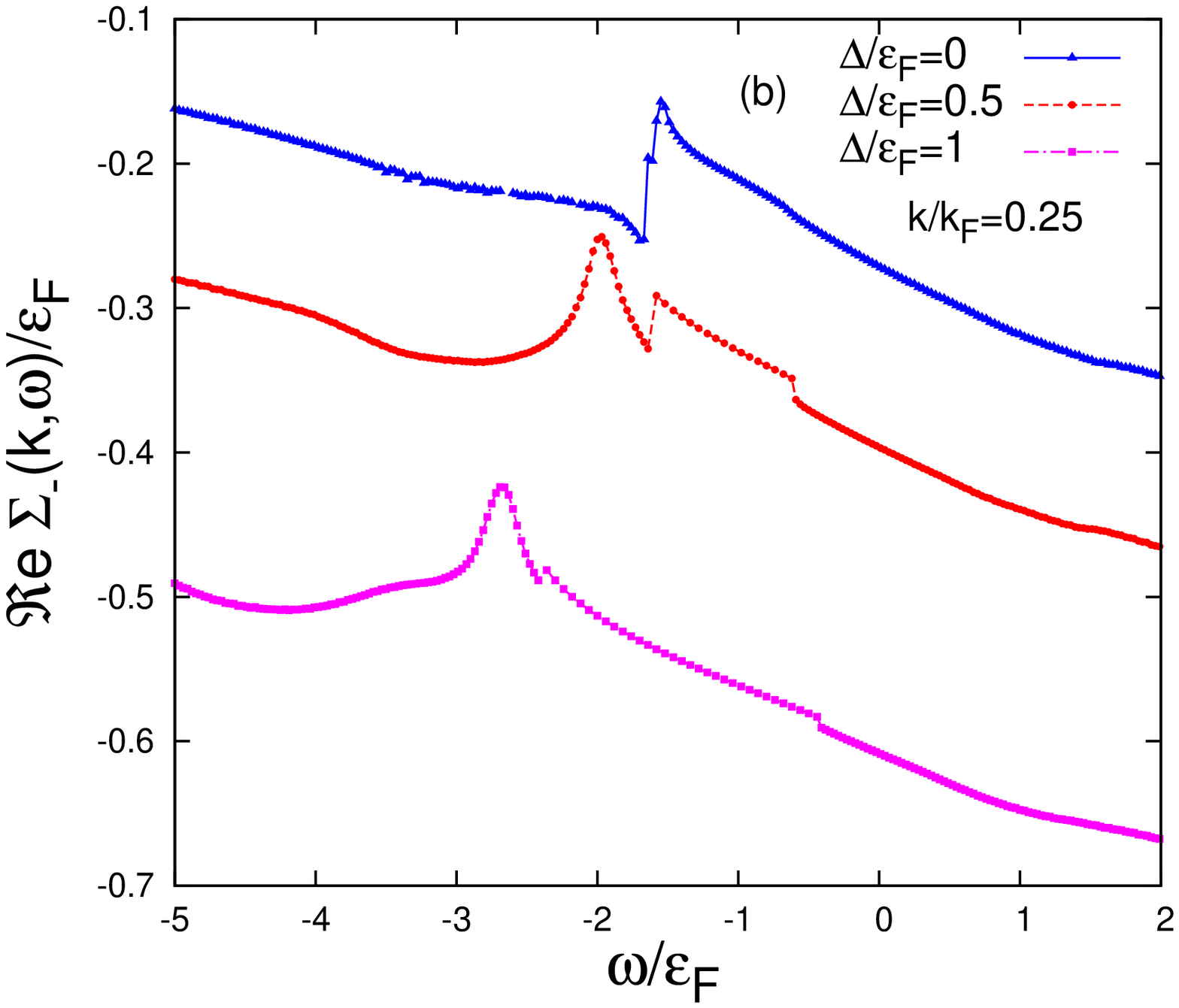}
\includegraphics[width=5cm]{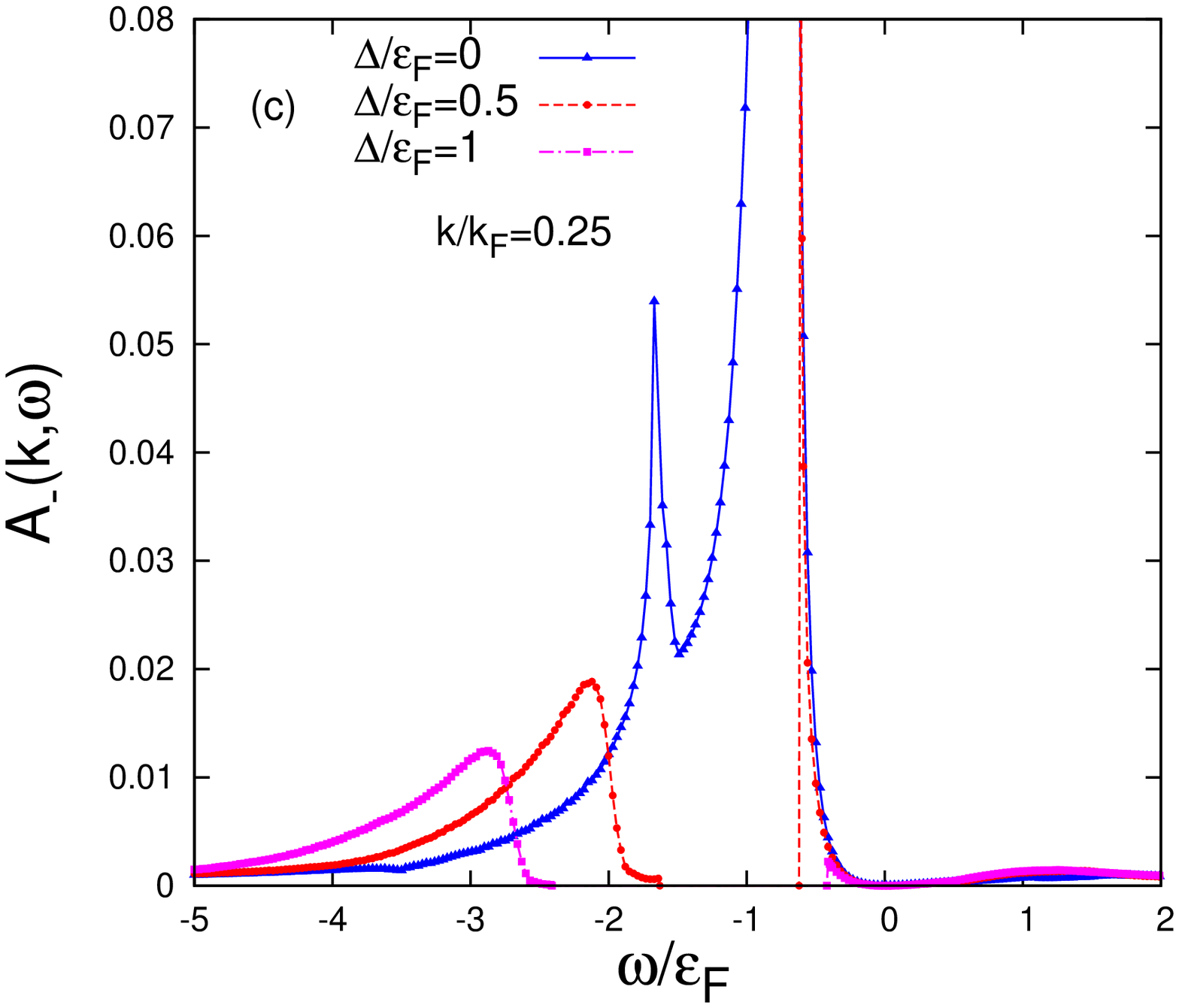}
\caption{(Color online) The imaginary (a) and real part (b) of the $(-)$ channel of retarded self-energy
as a function of $\omega$ for the various
energy gaps at $\Lambda=100$. The self-energy are measured
from the interaction contribution of the chemical potential $\Re e
\Sigma_-(k_{\rm F},\omega=0)$. The QP spectral function as a
function of $\omega$ (c), for the various energy
gaps at $\Lambda=100$.} \label{}
\end{figure}

The valance band self-energy contributions are shown in Fig.~8. There is an area of frequencies is associated to the gap value in which no QP could exist in $\Im m \Sigma_{-}$ exactly the same as the conduction band. The $s=+1$ and $-1$ peaks in $\Im m \Sigma_s$ in Figs.~4 and 8 separate at finite $k$ because of chirality factors which emphasize ${\bf k}$ and ${\bf q}$ in nearly parallel directions for conduction band and ${\bf k}$ and ${\bf q}$ in nearly opposite directions for valance band states. Consequently, at finite $\Delta$, the QP peak of $A_{-}(k,\omega)$ which is broaden shifts toward the left in the opposite behavior of $A_{+}(k,\omega)$. These feature have significant effects in the interacting electron density of states.

\begin{figure}[h]
\includegraphics[width=7cm]{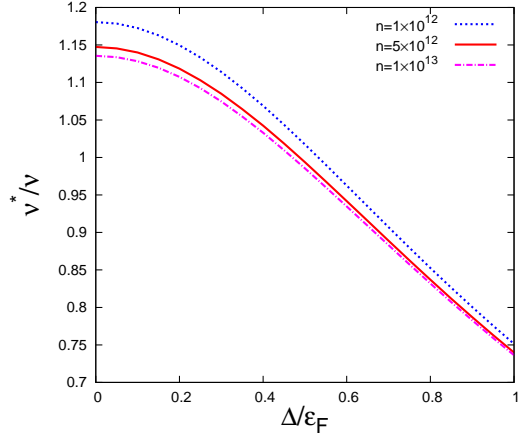}
\caption{(Color online) The renormalized velocity (+ channel) as a function of
the energy gap for various densities at $\alpha_{gr}=1$.}
\label{}
\end{figure}

It is essential to note that the satellite band which is theoretically predicted~\cite{im1}
for gapless garphene has not been seen in experiments. There are several reasons that could wash
out this feature. For example, the plasmon damping, disorder effects, electron interactions with the buffer layer and importantly the effect of gap at Dirac point.

One of the important information which can be extracted from ARPES
spectra is the renormalized Fermi velocity $v^*$. A consequence of the interaction is a Fermi velocity renormalization from the backflow of the fluid around a moving particle. The density of states at the Fermi energy is also changed. The QP
energy measured from the chemical potential of interacting system
$\delta\varepsilon_{sk}^{QP}$, can be calculated by solving self
consistently the Dyson equation $\delta\varepsilon_{s{\bf
k}}^{QP}=\xi_{s{\bf k}}+\Re e[\delta \Sigma_s^{\rm ret}({\bf {\bf
k}},\omega)]|_{\omega=\delta\varepsilon_{s{\bf
k}}^{QP}/\hbar}$.\cite{Giuliani} In the isotropic systems the
QP energy, depends on the magnitude of ${\bf k}$.
Expanding $\delta\varepsilon_{sk}^{QP}$ to first order in $k-k_{\rm F}$
we can write $\delta\varepsilon_{sk}^{QP}\simeq\hbar v^*_s(k-k_{\rm F})$
which effectively defines the renormalized velocity as $\hbar
v^*_s={d\delta\varepsilon_{sk}^{QP}}/{dk}|_{k=k_{\rm F}}$. From the
Dyson equation we can calculated the renormalized Fermi velocity
as~\cite{fateme,velocity1,velocity2}
\begin{eqnarray}
\frac{v^*_s}{v_s}=\frac{\varepsilon_{\rm F}E_{\rm F}^{-1}+(\hbar
v_s)^{-1}\partial_k\Re e[\delta\Sigma_s^{\rm ret}({\bf
k},\omega)]|_{\omega=0,k=k_{\rm F}}}{1-\hbar^{-1}\partial_{\omega}\Re
e[\delta\Sigma_s^{\rm ret}({\bf k},\omega)]|_{\omega=0,k=k_{\rm F}}},
\end{eqnarray}
where $v_s=sv_{\rm F}$. It is found before~\cite{fateme,velocity1,velocity2,gw1,gw2} that
electron-electron interaction increases the renormalized Fermi
velocity in gapless graphene sheets which this behavior is in
contrast to conventional 2DES.~\cite{asgari1,asgari}

Fig.~9 shows the
renormalized Fermi velocity in unit of the bare Fermi velocity as
a function of band gap for various carrier densities. The renormalized Fermi velocity decreasing with increasing the gap value. $v^*$ is density independent after $\Delta=0.8 \varepsilon_{\rm F}$ which is in good agreement with recent experiment observation.~\cite{dop_lanzara1,dop_lanzara2}

Finally we calculated a band gap renormalization (BGR).~\cite{bgr1,bgr2,bgr3,bgr4} The
BGR for conductance band is given by the QP self-energy
at the band edge, namely $\rm BGR=\Re e\Sigma_+^{\rm ret}({\bf
k}=0,\omega=(\Delta-E_{\rm F})/\hbar)$. Fig.~10 shows the BGR
for the various gap values as a function of the electron density. The BGR decreases by increasing of the
electron density and in the small
energy gap values, it is less density dependent respect to large
energy gap values. In gapless case, we have
obtained a induced band gap or kink due to many-body
electron-electron interactions and it tends to a constant with
increasing the electron density.~\cite{im1,im2,gw1,gw2} This feature is in agreement with the results obtained within {\it ab intio} DFT calculation.~\cite{gw1}
\begin{figure}[h]
\includegraphics[width=7cm]{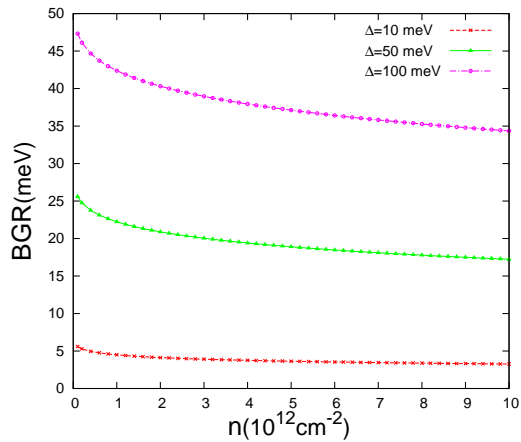}
\caption{(Color online) The band gap renormalization as a function
of electron density for various energy gap at $\alpha_{gr}=1$.}
\label{}
\end{figure}

\section{SUMMERY AND CONCLUSION}

We have revisited the problem of the microscopic
calculation of the QP self-energy and many-body effective velocity
suppression in a gapped graphene when the conduction band is partially occupied. We have performed a  systematic study is based on the many-body G$_0$W approach that is established upon the random-phase-approximation and on graphene's
massive Dirac equation continuum model. We have carried out extensive calculations of both the real and the imaginary part of the
QP self-energy and discussed about the interband and intraband contributions in the scattering process in the presence of gap value. We have also presented results for the effective velocity
and for the band gap renormalization over a wide range of coupling strength.
Accordingly, we have critically examined the merits
of the gap values in dynamical QP properties.

Most feature of mass generating in graphene
is the washing out of the plasmaron peak in the spectral weight.
Increasing of the gap value makes density independent behavior
of the renormalized Fermi velocity. We have shown that
the band gap renormalization in gapped graphene decreases by
increasing the carrier density at large $\Delta$. This is in contrast with the gapless case in which
many body electron-plasmon interactions induce a very small gap in band
structure. These distinct features of the massive Dirac's Fermions
are related to mixing of the chiralities and reduce of the
interband transitions in graphene sheets.

\textbf{Acknowledgment}

R. A. would like to thank the Scuola Normale Superiore, Pisa, Italy for its hospitality during the period when the final stage of this work was carried out. A. Q. supported by IPM grant.


\begin{thebibliography}{99}
\bibitem{novoselov1}
    K. S. Novoselov, A. K. Geim, S. V. Morozov, D. Jiang,
    Y. Zhang, S. V. Dubonos, I. V. Grigorieva, A. A. Firsov 2004 \textit{Science} {\bf 306} 666~.
\bibitem{novoselov2}
    K. S. Novoselov, A. K. Geim, S. V. Morozov, D. Jiang, M. I. Katsnelson, I.
    V. Grigorieva, S. V. Dubonos and A. A. Firsov 2005 \textit{Nature} {\bf 438} 197~.
    \bibitem{novoselov3}
    K. S. Novoselov, D. Jiang, F. Schedin, T. J. Booth, V. V. Khotkevich, S.
    V. Morozov, and A. K. Geim 2005 \textit{Proc. Nat. Acad. Sci.} {\bf 102} 10451~.
    \bibitem{novoselov4}
    Y. Zhang, Y. Tan, H. L. Stormer, P. Kim 2005 \textit{Nature} \textbf{438} 201~.
\bibitem{carbon}
    P. Avouris, Z. Chen, and V. Perebeinos, 2007 \textit{Nature Nanotech.} \textbf{2} 605~.
\bibitem{graphene_rev1}
    A. H. Castro Neto, F. Guinea, N. M. Peres, K. S. Novoselov, and A. K. Geim 2009 \textit{Rev. Mod. Phys.} {\bf 81} 109~.
    \bibitem{graphene_rev2}
    C. W. Beenakker 2008 \textit{Rev. Mod. Phys.} {\bf 80} 1337~.
    \bibitem{graphene_rev3}
    A. K. Geim and P. Kim, 2008 \textit{Sci. Am.} \textbf{298} 90~.
    \bibitem{graphene_rev4}
    A. K. Geim and A. H. MacDonald 2007 \textit{Physics Today} \textbf{60} 35~.
    \bibitem{graphene_rev5}
    A. K. Geim and K. S. Novoselov 2007 \textit{Nature Mater.} \textbf{6} 183~.
    \bibitem{graphene_rev6}
    M. I. Katsnelson, 2007 \textit{Materials Today} \textbf{10} 20~.
\bibitem{tomadin}
    M. Polini, A. Tomadin, R. Asgari and A. H. MacDonald 2008 \textit{Phys. Rev. B} {\bf 78} 115426~.
\bibitem{thermal}
    A. A. Balandin, S. Ghosh, W. Bao, I. Calizo, D. Teweldebrhan, F. Miao, C. N.
    Lau 2008 \textit{Nano Lett.} {\bf 8} 902~.
\bibitem{morozov1}
    S.V. Morozov, K.S. Novoselov, M.I. Katsnelson, F. Schedin, D.C. Elias, J.A. Jaszczak, A.K.
    Geim 2008 \textit{Phys. Rev. Lett.} {\bf 100} 016602~.
\bibitem{morozov2}
    K. I. Bolotin, K. J. Sikes, J. Hone, H. L. Stormer, and P.
    Kim 2008 \textit{Phys, Rev, Lett.} {\bf 101} 096802~.
    \bibitem{morozov3}
    Xu Du, Ivan Skachko, Anthoy Barker and Eva Y. Andrei 2008 \textit{Nature Nanotech.} {\bf 3} 491~.
    \bibitem{morozov4}
    K.I. Bolotin, K.J. Sikes, Z. Jiang, M. Klima, G. Fudenberg, J. Hone, P. Kim, H.L.
    Stormer 2008 \textit{Solid State Commun.} {\bf 146} 351~.
\bibitem{eng1}
    K. Eng, R. N. McFarland, and B. E. Kane 2005 \textit{Appl. Phys. Lett.} {\bf 87} 052106~.
    \bibitem{eng2}
    E. H. Hwang and S. Das Sarma 2007 \textit{Phys. Rev. B} {\bf 75} 073301~.
\bibitem{strong1}
    C. Lee, X. Wei, J. W. Kysar, J. Hone 2008 \textit{Science} \textbf{321} 385~.
    \bibitem{strong2}
    M. Neek-Amal and R. Asgari 2009 arXiv:0903.5035
\bibitem{dop1}
    T. O. Wehling, K. S. Novoselov, S. V. Morozov, E. E. Vdovin, M. I. Katsnelson, A. K. Geim, A. I.
    Lichtenstein 2008 \textit{Nano Lett.} {\bf 8} 173~.
    \bibitem{dop2}
    I. Gierz, C. Riedl, U. Starke, C. R. Ast, K. Kern 2008 \textit{Nano Lett.} {\bf 8} 4603~.
\bibitem{dop_lanzara1}
    S. Y. Zhou, D. A. Siegel, A. V. Fedorov, and A. Lanzara 2008 \textit{Phys. Rev. Lett.} {\bf 101} 086402~.
 \bibitem{dop_lanzara2}
    D. A. Siegel, S. Y. Zhou, F. El Gabaly, A. V. Fedorov, A. K. Schmid, and
    A. Lanzara, 2008 \textit{Appl. Phys. Lett.} \textbf{93} 243119~.
\bibitem{FET1}
    Y. Lin, K. A. Jenkins, A. Valdes-Garcia, J. P. Small, D. B. Farmer, and P. Avouris
    2009 \textit{Nano Lett.} \textbf{9} 422~.
    \bibitem{FET2}
    J. Kedzierski, P. Hsu, P. Healey, P. W. Wyatt, C. L. Keast, M.
    Sprinkle, C. Berger, and W. A. de Heer 2008 \textit{IEEE Trans. Electron Devices} \textbf{55} 2078~.
\bibitem{gap}
    K. Novoselov 2007 \textit{Nature Mater.} \textbf{6} 720~.
\bibitem{gap_ribbon1}
    Y. W. Son, M. L. Cohen and S. G. Louie 2006 \textit{Phys. Rev. Lett.} {\bf 97} 216803~.
    \bibitem{gap_ribbon2}
    M. Y. Han, B. Ozyilmaz, Y. Zhang and P. Kim 2007 {\it Phys. Rev. Lett.} {\bf 98} 206805~.
    \bibitem{gap_ribbon3}
    Li Yang, Cheol-Hwan Park, Young-Woo Son, M. L. Cohen, and S. G. Louie 2007 \textit{Phys. Rev. Lett.} \textbf{99} 186801~.
    \bibitem{gap_ribbon4}
    D. Finkenstadt, G. Pennington, and M. J. Mehl 2007 \textit{Phys. Rev. B} \textbf{76} 121405(R)~.
    \bibitem{gap_ribbon5}
    Y.-W. Son, M. L. Cohen and S. G. Louie 2006 \textit{Nature} {\bf 444} 347~.
\bibitem{gap_spin1}
    Xue-Feng Wang and T. Chakraborty 2007 \textit{Phys. Rev. B} \textbf{75} 033408~.
    \bibitem{gap_spin2}
    Y. Yao, F. Ye, X. L. Qi, S. C. Zhang, and Z. Fang 2007 \textit{Phys. Rev. B} \textbf{75} 041401(R)~.
    \bibitem{gap_spin3}
    C. L. Kane and E. J. Mele 2005 \textit{Phys. Rev. Lett.} \textbf{95} 226801~.
    \bibitem{gap_spin4}
    H. Min, J. E. Hill, N. A. Sinitsyn, B. R. Sahu, L. Kleinman, and A. H. MacDonald 2006 \textit{Phys. Rev. B} \textbf{74} 165310~.
\bibitem{gapsub1}
    G. W. Semenoff 1984 \textit{Phys. Rev. Lett.} \textbf{53} 2449~.
    \bibitem{gapsub2}
    K. Ziegler, 1996 \textit{Phys. Rev. B} \textbf{53} 9653~.
    \bibitem{gapsub3}
    V. P. Gusynin, S. G. Sharapov, J. P. Carbotte 2007 \textit{Int. J. Mod. Phys. B} \textbf{21} 4611~.
    \bibitem{gapsub4}
    A. Bostowick, T. Ohta, J. L. McCesney, K. V. Emtsev, T.
    Seyller, K. Horn and E. Rotenberg 2007 \textit{New J. Phys.} \textbf{9} 385~.
\bibitem{kekule1}
    C.-Y. Hou, C. Chamon, and C. Mudry 2007 \textit{Phys. Rev. Lett.} \textbf{98} 186809~.
\bibitem{lanzara1}
    S. Y. Zhou, G. H. Gweon, A. V. Federov, P. N. First, W. A. de
    Heer, D. H. Lee, F. Guinea, A. H. Castro Neto, and A. Lanzara 2007 \textit{Nature Mater.} {\bf 6} 770~.
    \bibitem{lanzara2}
    S.Y. Zhou, D.A. Siegel, A.V. Fedorov, and A. Lanzara 2008 \textit{Physica E} {\bf 40}, 2642~.
\bibitem{gruneis1}
    A. Gr\"{u}neis and D. V. Vyalikh 2008 \textit{Phys. Rev. B} {\bf 77} 193401
\bibitem{gruneis2}
    A. Gr\"{u}neis, K. Kummer and D. V. Vyalikh 2009 \textit{New J. Phys.} {\bf 11} 073050~.
\bibitem{eva}
    G. Li, A. Luican, and E. Y. Andrei 2009 \textit{Phys. Rev. Lett.} {\bf 102} 176804~.
\bibitem{giovannetti}
    G. Giovannetti, P. A. Khomyako, G. Brocks, P. J. Kelly and J. Van den Brink 2007 \textit{Phys. Rev. B} {\bf 76} 073103~.
\bibitem{im1}
    M. Polini, R. Asgari, G. Borghi, Y. Barlas, T. Pereg-Barnea, and A. H. MacDonald 2008 \textit{Phys. Rev. B} {\bf 77} 081411(R)~.
\bibitem{alireza}
    A. Qaiumzadeh and R. Asgari 2009 \textit{Phys. Rev. B} {\bf 79} 075414~.
\bibitem{ribeiro1}
    R. M. Ribeiro, N. M. R. Peres, J. Coutinho and P. R. Briddon, 2008 \textit{Phys. Rev. B} {\bf 78} 075442~.
\bibitem{ribeiro2}
    Eduardo V. Castro, K. S. Novoselov, S. V. Morozov, N. M. R. Peres, J. M. B. Lopes dos Santos, Johan Nilsson,
    F. Guinea, A. K. Geim and A. H. Castro Neto 2007 \textit{Phys. Rev. Lett.} {\bf 99} 216802~.
\bibitem{zanella}
    I. Zanella, S. Guerini, S. B. Fagan, J. Mendes Filho and A. G. Souza Filho 2008 \textit{Phy. Rev. B} {\bf 77} 073404~.
\bibitem{kruczynski}
    M. Mucha-Kruczy\'{n}ski, O. Tsyplyatyev, A. Grishin, E. McCann, Vladimir I. Fal'ko, Aaron Bostwick and Eli Rotenberg 2008 \textit{Phys. Rev. B} {\bf 77} 195403~.
\bibitem{kim}
    S. Kim, J. Ihm, H. J. Choi, and Y. Son 2008 \textit{Phys. Rev. Lett.} \textbf{100} 176802~.
\bibitem{antidot}
    T. G. Pedersen, A. Jauho, and K. Pedersen 2009 \textit{Phy. Rev. B} \textbf{79} 113406~.
\bibitem{Giuliani}
    G. F. Giuliani and G. Vignale 2005 \textit{Quantum Theory of The Electron
    Liquid} (Cambridge University Press, Cambridge, England)~.
\bibitem{yafis}
    Y. Barlas, T. Pereg-Barnea, M. Polini, R. Asgari and A. H.
    MacDonald 2007 \textit{Phys. Rev. Lett.} {\bf 98} 236601~.
\bibitem{pyatkovskiy1}
    P. K. Pyatkovskiy, 2009 \textit{J. Phys.: Condens. Matter } {\bf 21} 025506~.
    \bibitem{pyatkovskiy2}
    B. Wunsch, T. Stauber, F. Sols and F. Guinea 2006 \textit{New J. Phys.} \textbf{8} 318
\bibitem{kaminski1}
    A. Kaminski and H. M. Fretwell 2005 \textit{New J. Phys.} \textbf{7} 98~.
    \bibitem{kaminski2}
    A. Damascelli, Z. Hussain, and Z.-X. Shen 2003 \textit{Rev. Mod. Phys.} {\bf 75} 473~.
\bibitem{fateme}
    A. Qaiumzadeh, F. K. Joibari, and R. Asgari 2008 arXv: 0810.4681~.
\bibitem{inelastic1}
    E. H. Hwang, BenYu-Kaung Hu, and S. Das Sarma 2007 \textit{Phys. Rev. B} {\bf 76}115434~.
    \bibitem{inelastic2}
    W. Tse, E. H. Hwang, and S. Das Sarma 2008 \textit{Appl. Phys. Lett.} \textbf{93} 023128~.
\bibitem{asgari1}
    R. Asgari, B. Davoudi, M. Polini, G. F. Giuliani, M. P.
    Tosi, and G. Vignale 2005 \textit{Phys. Rev. B} {\bf 71} 045323~.
    \bibitem{asgari2}
    G. F. Giuliani and J. J. Quinn 1982 \textit{Phys. Rev. B} {\bf 26} 4421~.
\bibitem{im2}
    E. H. Hwang and S. Das Sarma 2008 \textit{Phys. Rev. B} {\bf 77} 081412(R)~.
\bibitem{velocity1}
    M. Polini, R. Asgari, Y. Barlas, T. Pereg-Barnea, A. H.
    MacDonald 2007 \textit{Solid State Commun.} {\bf 143} 58~.
    \bibitem{velocity2}
    A. Qaiumzadeh, N. Arabchi, R. Asgari 2008 \textit{Solid State Commun.} {\bf 147} 172
\bibitem{gw1}
    Paolo E. Trevisanutto, Christine Giorgetti, Lucia Reining, Massimo Ladisa and Valerio Olevano 2008
    \textit{Phys. Rev. Lett.} \textbf{101} 226405~.
\bibitem{gw2}
    C. Park, F. Giustino, C. D. Spataru, M. L. Cohen, and S. G.
    Louie 2009 \textit{Phys. Rev. Lett.} \textbf{102} 076803~.
\bibitem{asgari}
    R. Asgari and B. Tanatar 2006 {\it Phys. Rev. B} {\bf 74} 075301~.
\bibitem{bgr1}
    Y. Zhang and S. Das Sarma 2005 \textit{Phys. Rev. B} {\bf 72} 125303~.
    \bibitem{bgr2}
    S. Das Sarma, R. Jalabert, and S. R. Eric Yang 1990 \textit{Phys. Rev. B} {\bf 41} 8288~.
\bibitem{bgr3}
    K. F. Berggren and B. E. Sernelius 1984 \textit{Phys. Rev. B} {\bf 29} 5575~.
\bibitem{bgr4}
    K. F. Berggren and B. E. Sernelius 1981 \textit{Phys. Rev. B} {\bf 24} 1971~.


\end{thebibliography}
\end{document}